\newif\ifarxiv
\arxivtrue

\documentclass{article}

\usepackage[round]{natbib}  %
\bibpunct{(}{)}{;}{a}{}{,}  %
\usepackage{hyperref}       %
\usepackage{url}            %
\usepackage{booktabs}       %
\usepackage{amsfonts}       %
\usepackage{amsmath,amssymb}
\usepackage{nicefrac}       %
\usepackage{multirow}
\usepackage{bm}
\newcommand{\matr}[1]{\bm{#1}}
\usepackage{multicol}
\usepackage{subcaption}
\usepackage[title]{appendix}%

\ifarxiv
    \usepackage{graphicx}
    \usepackage{xcolor}
    \usepackage{PRIMEarxiv}
    \newcommand{\orcid}[1]{}         %
\fi

\begin{document}

\ifarxiv
    \title{Spatially heterogeneous power-law attenuation with multiple relaxation mechanisms for ultrasound modeling}
    \author{Masashi Sode$^{1,2}$\quad Gianmarco Pinton$^{1,2,\ast}$}
    \maketitle
    \vspace{-0.25in}
    {\centering\small
        $^{1}$Lampe Joint Department of Biomedical Engineering, University of North Carolina at Chapel Hill, Chapel Hill, NC, United States\par
        $^{2}$Lampe Joint Department of Biomedical Engineering, North Carolina State University, Raleigh, NC, United States\par
        $^{\ast}$Corresponding author: gia@email.unc.edu\par
    }
    \vspace{0.15in}
\else
    \articletype{Paper} %

    \title{Spatially heterogeneous power-law attenuation with multiple relaxation mechanisms for ultrasound modeling}
    \author{Masashi Sode$^{1, 2}$\orcid{0000-0002-3685-7378}, and Gianmarco Pinton$^{1,2,*}$\orcid{0000-0002-4896-1439}}
    \affil{$^1$Lampe Joint Department of Biomedical Engineering, University of North Carolina at Chapel Hill, Chapel Hill, NC, United States}
    \affil{$^2$Lampe Joint Department of Biomedical Engineering, North Carolina State University, Raleigh, NC, United States}
    \affil{$^*$Author to whom any correspondence should be addressed.}
    \email{gia@email.unc.edu}
\fi

\ifarxiv\else
\fi
\begin{abstract}
    \textbf{Objective} The soft tissue attenuation laws have a magnitude and frequency dependence that varies across tissue types and generally follow power laws. An accurate model of ultrasound propagation in the human body thus may require spatially heterogeneous power-law attenuation $\alpha(\mathbf{x},f) = \alpha_0(\mathbf{x}) f^{y(\mathbf{x})}$. However, a spatially heterogeneous representation of frequency-dependent attenuation is technically challenging, so existing methods introduce simplifying assumptions. For example, prior approaches such as Fullwave 2 achieved $<5\%$ error for individual tissue types but required manual parameter tuning for each $(\alpha_0, y)$ pair, limiting the construction of realistic tissue libraries.\\
    \textbf{Approach} We introduce a calibration framework that uses derivative-free optimization to systematically fit relaxation parameters across diverse tissue combinations spanning $\alpha_0 = 0.0022$--$1.0$ dB/(MHz$^y$ cm) and $y = 0.4$--$2.0$. The Nelder-Mead algorithm minimizes complex-wavenumber mismatch. The attenuation is extended to a convolutional perfectly matched layer, where the same relaxation formulation is used in the boundaries.\\
    \textbf{Main results} The method achieves mean errors below $3\%$ over $1$--$20$ MHz with dispersion error of $1.1 \pm 0.8$ m/s across the clinically relevant core region ($y = 0.7$--$1.4$). Boundary reflections remain below $-50$ dB for clinically relevant tissue exponents ($y \leq 1.5$). We validated the method with two-layer muscle/fat/liver models and confirmed per-layer accuracy ($<2.5\%$ normalized error). A 3D abdominal simulation using the Visible Human dataset demonstrates stable propagation with voxel-level heterogeneity in both $\alpha_0(\mathbf{x})$ and $y(\mathbf{x})$.\\
    \textbf{Significance} The open-source multi-GPU implementation (Fullwave 2.5) provides a practical tool for patient-specific therapy planning, training data generation, estimation of acoustic radiation force, quantitative imaging, and inverse problem applications.
\end{abstract}
 \ifarxiv
\fi

\section{Introduction}

Attenuation affects wave propagation in multiple ways
including frequency-dependent amplitude decay~\citep{Szabo1994-eb,Holm2019-zd}, phase velocity altering through causality~\citep{Waters2000-bx}, and sound speed variation due to the pressure dependent
nonlinearity~\citep{huijssen2010iterative, Humphrey2000-vg}.
Accurate ultrasound simulation requires modeling these effects to predict the acoustic field in heterogeneous biological tissues for various applications, including diagnostic imaging~\citep{Szabo2013-bt}, tissue harmonic imaging~\citep{Tranquart1999-zf}, and the design of transducers and beamforming algorithms~\citep{Pajek2012-xn, Jones2021-bd}. It is also necessary for quantitative ultrasound~\citep{Mamou2013-gc}, and the estimation of therapeutic doses in neuromodulation~\citep{Blackmore2019-wc,Darmani2022-ru, Matt2024-vd} and high-intensity focused ultrasound (HIFU)~\citep{Izadifar2020-jq}.

\subsection{Challenges of Power-Law Modeling}

Despite its importance, achieving high-fidelity modeling of attenuation has been limited due to the challenges of power-law modeling.
The attenuation in biological tissue often follows a power-law $\alpha(f) = \alpha_0 f^y$ with attenuation coefficient $\alpha_0$ and non-integer exponent $y$~\citep{Szabo1994-eb, Holm2019-zd}. The attenuation coefficient $\alpha_0$ and exponent $y$ vary significantly across different tissue types and anatomical locations~\citep{Wells1975-lq}.
For example, in soft tissues, the attenuation coefficient $\alpha_0$ typically ranges from 0.1 to 1.0 [dB/(MHz$^y$ cm)], while the exponent $y$ can vary from 0.5 to 2.0~\citep{Wells1975-lq}.
This variability necessitates flexible modeling approaches that can accurately capture the spatial heterogeneity of attenuation properties in biological tissues.

Power law modeling is challenging due to the dispersive nature of biological tissues and non-local in time behavior~\citep{Holm2019-zd}.
Dispersive media exhibit frequency-dependent phase velocity, meaning that different frequency components of the wave travel at different speeds.
It is a fundamental property of causal systems that attenuation and dispersion are intrinsically linked through the Kramers-Kr\"{o}nig relations~\citep{Waters2000-bx}.
The non-local in time behavior arises from the non-integer power-law exponent, which leads to a dependence of the current wavefield state on its entire history~\citep{Holm2019-zd}.
This non-local in time behavior poses challenges for numerical modeling, as it requires specialized memory techniques to accurately capture the time-dependent behavior of attenuation and dispersion.
Heterogeneous power-law exponent modeling is still unexplored in ultrasound simulations. Real tissues such as tumor margins and regions with fatty infiltration exhibit gradual transitions in the exponent $y$ at sub-millimeter scales. Fixed-exponent models force binary segmentation of such regions and lose realistic tissue heterogeneity. This limitation motivates the development of methods that support spatially varying exponents $y(\mathbf{x})$.
\subsection{Connection between relaxation spectra, power-law attenuation, and fractional models} \label{subsec:theoretical_framing}

The frequency response of a finite relaxation model can be written as a finite sum of Debye terms. In generic form, the dynamic compressibility (or equivalently the complex modulus/compliance representation) is
\begin{align}
    \chi(\omega) = \chi_\infty + \sum_{m=1}^{M} \frac{a_m}{\Omega_m + i\omega},
\end{align}
which is a rational function of $i\omega$ with a finite set of poles at $-\Omega_m$.
This finite-pole structure is fundamentally different from ideal broadband power-law attenuation, which is associated with fractional-order behavior and therefore non-rational frequency dependence.
Consequently, a finite number of relaxation mechanisms cannot exactly reproduce a power law over the full frequency axis; it can only approximate it over a bounded design band.

A continuum limit is obtained by replacing the discrete mechanisms with a nonnegative relaxation density:
\begin{align}
    \chi(\omega) = \chi_\infty + \int_0^\infty \frac{H(\Omega)}{\Omega + i\omega} \, d\Omega.
    \label{eq:continuous_relaxation_spectrum}
\end{align}
For power-law-type spectra $H(\Omega)$, this representation yields attenuation and dispersion proportional to fractional powers of frequency and is equivalent to fractional constitutive laws~\citep{Nasholm2011-xx,Holm2019-zd}.

In this perspective, approximating $\omega^y$ attenuation with finite relaxations is equivalent to approximating $(i\omega)^y$ using a rational function.
Because $(i\omega)^y$ introduces a branch cut in the complex-frequency plane, whereas finite relaxation models provide only pole expansions, accurate broadband approximation requires increasingly dense pole placement over the band of interest.
This is the source of slow convergence and conditioning difficulties in constrained optimization.

The discrete relaxation model used here should be interpreted as a quadrature approximation of a continuous relaxation spectrum.
This interpretation turns parameter fitting from an ad hoc curve fit into a controlled approximation to a known continuum limit.
\begin{figure*}[t]
    \centering
    \includegraphics[width=\textwidth]{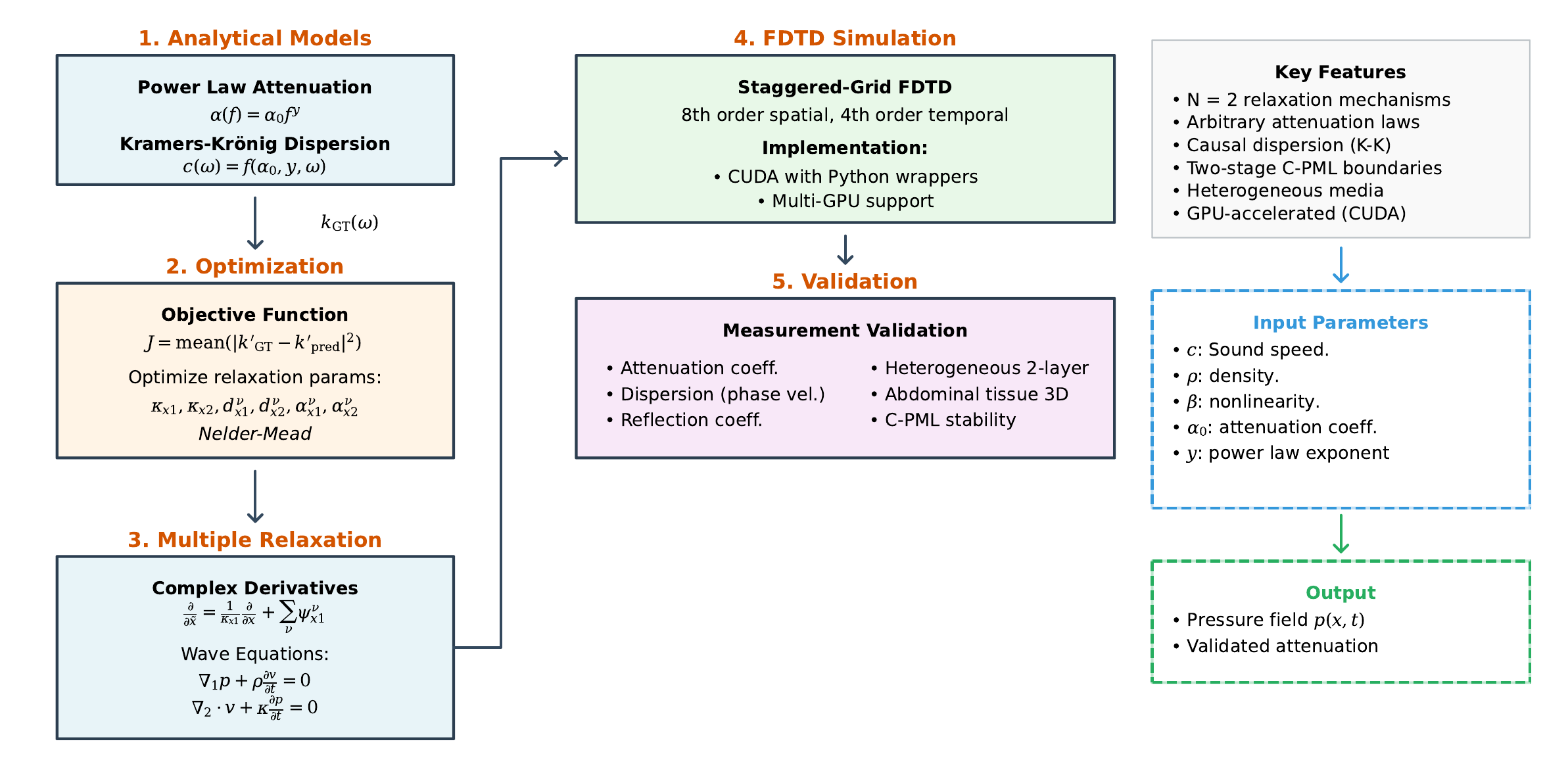}
    \caption{Overview of the Fullwave 2.5 workflow: (1) Analytical models define power-law attenuation and Kramers-Krönig dispersion. (2) Derivative-free optimization calibrates relaxation parameters by minimizing complex wavenumber mismatch. (3) Multiple relaxation mechanisms incorporate memory variables into wave equations. (4) GPU-accelerated FDTD simulation with 8th-order spatial accuracy. (5) Validation against analytical solutions for homogeneous, heterogeneous, and 3D scenarios.}
    \label{fig:method_overview}
\end{figure*}

\subsection{Related Work}

Many methods have been proposed to tackle the challenges of power-law attenuation modeling.
Mainly, they can be categorized into three groups: multiple relaxation models, space-fractional derivative based methods and time-fractional derivative based methods.

Firstly, multiple relaxation models have been proposed to approximate power-law attenuation by employing a set of auxiliary memory variables to capture the memory effects of the medium~\citep{Pierce2021-jz, Jimenez2015-rf, Pinton2009-uc, Sode2026-to}.
These models can effectively approximate power-law behavior by adjusting the parameters of the relaxation processes.
These models are often implemented in time-domain simulations using finite-difference time-domain (FDTD).
However, existing multiple relaxation models do not provide a heterogeneous power law attenuation model that can flexibly adjust the attenuation coefficient and exponent over a wide range.
For example, \citet{Sode2026-to} proposed a multiple relaxation model to approximate power-law attenuation in soft tissues, but their model is limited to a specific range of attenuation coefficients and exponents due to the manual parameter selection process.
\citet{Jimenez2015-rf} proposed a method to model power-law attenuation using multiple relaxation processes, but their approach is limited to homogeneous media and does not support heterogeneous attenuation profiles.

Secondly, time-fractional derivative based methods directly discretize wave equations containing fractional time derivatives such as formulations based on Szabo's wave equation~\citep{Szabo1994-eb, Chen2003-ro} or Caputo's wave equation~\citep{M-Caputo2011-jx}.
Standard methods of discretizing fractional time derivatives involve convolutions that require storing the entire wavefield history. This leads to high memory consumption and computational cost.
Recently, \citet{King2025-ym} proposed a static memory method to reduce the memory requirement. This method converts the time-fractional derivative into a sum of integrals that can be updated at each time step.

On the other hand, space-fractional derivative based methods model power-law attenuation using fractional spatial derivatives, such as the fractional Laplacian operator~\citep{Treeby2010-vh} including well-known k-Wave toolbox~\citep{Treeby2010-gb}.
This method avoids the need for storing the entire wavefield history by employing the fractional Laplacian operator instead of time-fractional derivatives.
Nevertheless, the operator is derived non-local in space, making it challenging to account for the local change of power law attenuation exponent in heterogeneous media.
Therefore, their implementation is limited in a homogeneous power-law attenuation exponent settings~\citep{Treeby2014-mk, Spa2025-kd}.

While time-fractional derivative based methods and space-fractional derivative based methods have been widely used for power-law attenuation modeling in ultrasound simulation, they face challenges in arbitrary attenuation law modeling, computational efficiency, and scalability for large-scale 3D simulations, due to the global nature of the Fourier transforms and the non-local nature of the fractional derivatives.

\subsection{Our Approach: Fullwave 2.5}

We introduce Fullwave 2.5, an automated calibration framework that extends the multiple relaxation approach of \citet{Sode2026-to}. While Fullwave 2 demonstrated $<5\%$ attenuation error over 1--20 MHz for specific tissue examples, it relied on manual grid search for parameter selection. This manual approach proved time-consuming: exploring different exponents required repeating the entire grid search from scratch, limiting the method to proof-of-concept demonstrations with carefully tuned parameters and creating a barrier to adoption for practical tissue libraries spanning diverse attenuation characteristics.

Fullwave 2.5 addresses this bottleneck through automated derivative-free optimization. The Nelder-Mead algorithm systematically calibrates relaxation parameters across 1328 $(\alpha_0, y)$ combinations spanning $\alpha_0 = 0.0022$--$1.0$ dB/(MHz$^y$ cm) and $y=0.4$--$2.0$, transforming the method from proof-of-concept into a practical deployment tool. This optimization is a one-time preprocessing step taking 10 hours that generates parameter lookup tables; the runtime simulation performance remains identical to Fullwave 2, as both use the same underlying FDTD implementation with C-PML formulation~\citep{Sode2026-to, Tan2014-qp, Komatitsch2007-et}. Crucially, automated parameter selection enables heterogeneous power-law exponent modeling with spatially varying $y(\mathbf{x})$, a capability that was computationally impractical with manual grid search. Table~\ref{tab:fullwave_comparison} summarizes the key differences between Fullwave 2 and Fullwave 2.5.

The framework allows researchers to generate parameter sets for arbitrary tissue libraries without expert manual tuning, making high-fidelity simulation accessible for applications requiring diverse tissue properties such as transcranial ultrasound, abdominal imaging, and quantitative ultrasound.

Figure~\ref{fig:method_overview} provides an overview of the complete Fullwave 2.5 workflow, illustrating the five main stages from analytical models through optimization, simulation, and validation. The optimization framework presented in Section~\ref{subsec:optimization_framework} describes the complex-wavenumber objective function and Nelder-Mead calibration process. Sections~\ref{subsec:heterogeneous_validation} and~\ref{subsec:3d_demo} demonstrate heterogeneous validation with spatially varying power-law exponents in both quantitative two-layer models and realistic 3D abdominal simulations.

We validate the proposed method against analytical solutions, demonstrating its accuracy and effectiveness in modeling frequency-dependent attenuation and dispersion in ultrasound simulations.

\begin{table}[t]
    \centering
    \caption{Comparison of Fullwave 2 and Fullwave 2.5 capabilities.}
    \label{tab:fullwave_comparison}
    \begin{tabular}{lcc}
        \hline
        Capability                           & Fullwave 2         & Fullwave 2.5            \\
        \hline
        Multiple relaxation FDTD             & \checkmark         & \checkmark              \\
        Parameter selection                  & Manual grid search & Automated optimization  \\
        Tissue library coverage              & 5 examples         & 1328 combinations       \\
        Heterogeneous $\alpha_0(\mathbf{x})$ & \checkmark         & \checkmark              \\
        Heterogeneous $y(\mathbf{x})$        & $\times$           & \checkmark (first time) \\
        Preprocessing time                   & Days per exponent  & 10 hours (full library) \\
        \hline
    \end{tabular}
\end{table}

\subsection{Contributions}
This work makes several key contributions to the modeling of frequency-dependent attenuation and dispersion in ultrasound simulations.
First, we introduce an automated derivative-free optimization framework, based on the Nelder-Mead algorithm, that systematically calibrates relaxation parameters across 1328 $(\alpha_0, y)$ tissue combinations. This framework replaces the manual grid search used in prior work and produces precise alignment between the analytical and modeled attenuation-dispersion characteristics over a broad range of clinically relevant tissue properties.

Building on this optimization pipeline, Fullwave 2.5 accurately reproduces power-law attenuation and its associated Kramers-Kr\"onig dispersion over $\alpha_0=0.0022$--$1.0$ dB/(MHz$^y$ cm) and $y=0.4$--$2.0$ in the $1$--$20$ MHz band, covering the full range of soft tissues encountered in diagnostic ultrasound. As a direct consequence, Fullwave 2.5 enables, to our knowledge for the first time, time-domain ultrasound simulation with spatially heterogeneous power-law exponents $y(\mathbf{x})$, allowing accurate modeling of realistic tissue distributions in which both the attenuation coefficient and the exponent vary at subwavelength scales.

Finally, Fullwave 2.5 is open-sourced and publicly available on GitHub\footnote{\url{https://github.com/pinton-lab/fullwave25}} with multi-GPU support for large-scale 3D simulations, and the optimized relaxation-parameter lookup tables for all $(\alpha_0, y)$ combinations are included in the repository for immediate use by the community.

\section{Background}

\subsection{Multiple relaxation framework from Fullwave 2} \label{sec:multiple_relaxation}
Fullwave 2.5 builds on the multiple relaxation framework of \citet{Sode2026-to}, which employs the convolutional perfectly matched layer (C-PML) formulation~\citep{Komatitsch2007-et} to model arbitrary attenuation laws in ultrasound. We summarize the key equations below, as the relaxation parameters are the optimization targets in this work.

The model uses a modified pressure-velocity equation with complex coordinate stretching:
\begin{align}
     & \nabla_1 p + \rho \cfrac{\partial \matr{v}}{\partial t} = 0         \label{eq:wave1} \\
     & \nabla_2 \cdot \matr{v} + \kappa \cfrac{\partial p}{\partial t} = 0 \label{eq:wave2}
\end{align}
where $p(\matr{x}, t)$ and $\matr{v}(\matr{x}, t)$ represent pressure and velocity wavefields, and $\rho(\matr{x})$ and $\kappa(\matr{x})$ denote density and compressibility.
The complex spatial derivatives $\nabla_1$ and $\nabla_2$ control attenuation and dispersion through auxiliary relaxation variables:
\begin{align}
    \cfrac{\partial}{\partial \tilde{x}} = \cfrac{1}{\kappa_{x1}} \cfrac{\partial}{\partial x} + \sum_{\nu=1}^{N} \psi_{x1} ^ {\nu} \label{eq:relaxation4}
\end{align}
The auxiliary relaxation variable $\psi_{x1}^{\nu}$ is calculated at each time step $n$ using the following recursive formula:
\begin{align}
    (\psi_{x1}^{\nu})^n = b_{x1}^{\nu} (\psi_{x1}^{\nu})^{n-1} + a_{x1}^{\nu} \left(\cfrac{\partial}{\partial x} \right)^{n-1/2} \label{eq:relaxation_psi}
\end{align}
where the coefficients $a_{x1}^{\nu}$ and $b_{x1}^{\nu}$ are defined as:
\begin{align}
    a_{x1}^{\nu} & = \cfrac{d_{x1}^{\nu}}{\kappa_{x1}(d_{x1}^{\nu} + \kappa_{x1} \alpha_{x1}^{\nu})} (b_{x1}^{\nu} - 1) \label{eq:relaxation_a} \\
    b_{x1}^{\nu} & = e^{-(d_{x1}^{\nu} / \kappa_{x1} + \alpha_{x1}^{\nu})\Delta t} \label{eq:relaxation_b}
\end{align}
Identical calculations apply for $\frac{\partial}{\partial \tilde{y}_i}$ and $\frac{\partial}{\partial \tilde{z}_i}$ in the $\nabla_1$ and $\nabla_2$ operators.
The relaxation parameters $\kappa_{xi}$, $d_{xi}^{\nu}$, and $\alpha_{xi}^{\nu}$ have the following physical interpretations: $\kappa_{xi}(\matr{x})$ represents linear scaling of the derivative (modifying wave velocity), $d_{xi}^{\nu}$ represents scaling-dependent damping, and $\alpha_{xi}^{\nu}$ represents scaling-independent damping.
For an isotropic relaxation model with $N$ mechanisms, there are $2 + 4N$ parameters. In this work, we use two relaxation mechanisms ($N=2$), yielding 10 parameters: $\kappa_{x1}$, $\kappa_{x2}$, $d^{1}_{x1}$, $d^{1}_{x2}$, $\alpha^{1}_{x1}$, $\alpha^{1}_{x2}$, $d^{2}_{x1}$, $d^{2}_{x2}$, $\alpha^{2}_{x1}$, and $\alpha^{2}_{x2}$ (where subscripts $x1$ and $x2$ correspond to $\nabla_1$ and $\nabla_2$).

For a complete derivation of the multiple relaxation model and C-PML formulation, see~\citep{Sode2026-to}.
\subsection{Dispersion relation}
The complex wavenumber $k$ quantifies dispersion and attenuation in this system:
\begin{align}
    k = \cfrac{\omega}{c} \left( \cfrac{1}{\kappa_{x1} \kappa_{x2}} - \cfrac{\gamma_1}{\kappa_{x2}} - \cfrac{\gamma_2}{\kappa_{x1}} + \gamma_1 \gamma_2 \right) ^ {-\frac{1}{2}} \label{eq:dispersion_relations}
\end{align}
where $c=1/\sqrt{\kappa \rho}$ and
\begin{align}
    \gamma_1 = \sum_{\nu=1}^{N} \cfrac{d^\nu_{x1}}{\kappa^2_{x1}} \left( \cfrac{1}{d_{x1}^\nu / \kappa_{x1} + \alpha_{x1}^{\nu} + i \omega}  \right) \label{eq:gamma1}
    \\
    \gamma_2 = \sum_{\nu=1}^{N} \cfrac{d^\nu_{x2}}{\kappa^2_{x2}} \left( \cfrac{1}{d_{x2}^\nu / \kappa_{x2} + \alpha_{x2}^{\nu} + i \omega}  \right)
\end{align}
This dispersion relation is used to define the optimization objective function in Section~\ref{subsec:optimization_framework}.

\section{Fullwave 2.5: Automated Optimization Framework}
We now describe the optimization and validation pipeline. This section presents the complete Fullwave 2.5 framework for modeling arbitrary power-law attenuation in ultrasound simulations. The approach integrates analytical target models, derivative-free optimization, multiple relaxation mechanisms, FDTD simulation, and comprehensive validation. Figure~\ref{fig:method_overview} illustrates the complete workflow and the relationships between these components.

\subsection{Analytical target models for attenuation and dispersion}
We employed the power law attenuation model to emulate the frequency-dependent attenuation behavior in biological tissues.
For dispersion, we utilized Kramers-Kr\"{o}nig relations to relate the attenuation and dispersion characteristics.
The power law attenuation model is defined as:
\begin{align}
    \alpha(f) = \alpha_0 f^{y} \label{eq:power_law_attenuation}
\end{align}
where $\alpha_0$ is the attenuation coefficient at a reference frequency, $f$ is the frequency, and $y$ is the power law exponent.
The Kramers-Kr\"{o}nig relations \citep{Waters2000-bx} provide a causal relationship between the real and imaginary parts of the complex wave number, which is defined as:
\begin{align}
    \cfrac{1}{c(\omega)} & = \cfrac{1}{c_0} + \alpha_0 \tan\left(\cfrac{\pi y}{2}\right) \left(\omega ^{y-1} - \omega_0^{y-1}\right), \quad y \neq 1 \label{eq:kk_relation}
\end{align}
where $c(\omega)$ is the frequency-dependent phase velocity, $c_0$ is the phase velocity at the reference frequency $\omega_0$, and $\alpha_0$ and $y$ are the parameters from the power law attenuation model.
The free parameters $\alpha_0$ and $y$ can be adjusted to fit the desired attenuation characteristics of the medium. For example, water has an attenuation coefficient of approximately $0.0022$ dB/cm/MHz$^2$ with a power law exponent of $2$, while soft tissues typically exhibit attenuation coefficients ranging from $0.1$ to $1.0$ dB/cm/MHz with power law exponents between $0.5$ and $1.5$.
This attenuation modeling and dispersion modeling are widely used in the ultrasound simulation research.
\subsection{Optimization framework for relaxation parameters} \label{subsec:optimization_framework}
This optimization framework addresses the manual grid-search bottleneck identified in Section~1.1, enabling systematic calibration of relaxation parameters across diverse tissue property combinations.
The objective function is defined using the complex wave number, which relates the dispersion and attenuation characteristics in the complex domain physically.
The objective function quantifies the difference between the analytical wave number, derived from the Kramers-Kr\"{o}nig relations (Equation \ref{eq:power_law_attenuation} and \ref{eq:kk_relation}), and the wave number predicted by the multiple relaxation model, as outlined in equation \ref{eq:dispersion_relations}.
The complex wavenumber-based objective function is defined as follows:
\begin{align}
    J(\matr{k}_{\text{GT}}, \matr{k}_{\text{pred}}) & = \text{mean}(|\matr{k}_{\text{GT}}' - \matr{k}_{\text{pred}}'|^2)
\end{align}
The dashed terms $\matr{k}'_{\text{GT}}$ and $\matr{k}'_{\text{pred}}$ represent the normalized complex wave numbers for the analytical and predicted models, respectively, defined as:
\begin{align}
    \matr{k}'_X & = (1-w) \cfrac{ \matr{k}_{X, \text{real}}}{\max(\matr{k}_{\text{GT}, \text{real}}) - \min(\matr{k}_{\text{GT}, \text{real}})}                                      \\ \nonumber
                & + i w \cfrac{\matr{k}_{X, \text{imag}}}{\max(\matr{k}_{\text{GT}, \text{imag}}) - \min(\matr{k}_{\text{GT}, \text{imag}})}, \quad X \in \{\text{GT}, \text{pred}\}
\end{align}
and the analytical wave number $\matr{k}_{\text{GT}}$ and the predicted wave number $\matr{k}_{\text{pred}}$ are defined as:
\begin{align}
    k_{\text{GT}} (\omega)   & = \cfrac{\omega}{c (\omega)} + i \alpha (\omega)                                                                                                           \\
    k_{\text{pred}} (\omega) & = \cfrac{\omega}{c}\left( \cfrac{1}{\kappa_1 \kappa_2} - \cfrac{\gamma_1}{\kappa_2} - \cfrac{\gamma_2}{\kappa_1} + \gamma_1 \gamma_2\right)^{-\frac{1}{2}}
\end{align}
This objective function calculates the mean squared error between the normalized real and imaginary parts of the analytical wave number $\matr{k}_{\text{GT}}$ and the predicted wave number $\matr{k}_{\text{pred}}$ over a range of frequencies.
The normalization is essential due to the significant order of magnitude difference between the real and imaginary parts of the complex wave number.
Thus, balancing their contributions in the loss function is crucial for effective optimization.
The weight parameter $w$ balances the contributions of the real and imaginary parts in the loss function.

In this study, we handle the dispersion as a regularization term and focus on optimizing the attenuation term. Importantly, the weight parameter $w$ is not a single fixed value but rather an adaptive free parameter that is individually selected for each $(\alpha_0, y)$ combination based on attenuation fitting performance. For each tissue parameter combination, we evaluate multiple weight values ranging from $0.1$ to $0.9$ and select the weight that yields the best attenuation accuracy.
We found that dispersion is less sensitive to the weight parameter compared to attenuation. Therefore, this approach allows us to prioritize the optimization of attenuation characteristics while maintaining a reasonable representation of dispersion.
In addition to the original objective above, we sample frequencies logarithmically over the target band so that fitting error is distributed more uniformly across decades.
This is equivalent to emphasizing log-frequency behavior of power-law responses and reduces over-emphasis of the highest sampled frequencies.

Physical admissibility is satisfied implicitly by the parameter ranges in Table~\ref{tab:optimized_relaxation_parameters}, which restrict each stretch $i\in\{1,2\}$ (corresponding to $\nabla_1$ and $\nabla_2$ in Section~\ref{sec:multiple_relaxation}) to nonnegative effective strengths and positive relaxation times:
\begin{align}
    \kappa_{xi}>0,\quad d_{xi}^{\nu}\ge 0,\quad \alpha_{xi}^{\nu}>0,\quad \nu=1,\dots,N.
\end{align}
The Nelder-Mead simplex operates on the log-scale-transformed $d_{xi}^{\nu}$ and $\alpha_{xi}^{\nu}$, so these inequalities are satisfied automatically and preserve passive and causal behavior in the equivalent Debye representation.

To minimize the objective function, we evaluated several optimization algorithms, including COBYQA \citep{rago_thesis}, Nelder-Mead method \citep{Gao2012-iv}, and COBYLA \citep{Powell1994-ts, Zhang_2023}, which are derivative-free optimization algorithms, as well as Adam \citep{Kingma2014-nx}, which is a gradient-based optimization algorithm.
Based on comparative optimization results presented in Section~\ref{subsec:optimization_comparison}, we adopted the adaptive Nelder-Mead algorithm for all relaxation parameter optimizations in this study, as it consistently outperformed other methods by 2--3 orders of magnitude in terms of the objective function value.
We used SciPy's \texttt{optimize.minimize} function \citep{2020SciPy-NMeth} with the Nelder-Mead method to perform the optimization.

The optimization algorithm iteratively updates the relaxation parameters to minimize the objective function until convergence is achieved or a maximum number of iterations is reached.
The termination criteria for each optimization algorithm are set based on the convergence behavior observed during preliminary experiments.
For COBYLA and COBYQA, we set the maximum number of iterations to $5 \times 10^4$ and the tolerance to $1 \times 10^{-25}$. The convergence criteria for the Nelder-Mead method were set with a maximum of $5 \times 10^4$ iterations and an absolute tolerance of $1 \times 10^{-25}$.
Each individual $(\alpha_0, y)$ pair converges in a few seconds, and the complete optimization pipeline for all 1328 tissue parameter combinations takes approximately 10 hours on a standard workstation (Intel Core i9-13900K, 32 cores, 5.8 GHz max frequency).

The variables $\alpha^\nu_{x}$ and $d^\nu_{x}$ are converted to logarithmic scale during the optimization to improve the convergence speed.
Table \ref{tab:optimized_relaxation_parameters} summarizes the variables used in the optimization process, including their ranges, scales, and initial values.
The initial values were selected based on preliminary experiments and provided consistent convergence across all tissue parameter combinations. While derivative-free optimization methods are generally sensitive to initialization, the selected initial values proved robust for the target parameter space.
\begin{table}[t]
    \centering
    \caption{Optimization variables for relaxation parameters. $\nu$ indicates the relaxation mechanism index, where $\nu = 1, 2$ for the two relaxation mechanisms used in this study.}
    \begin{tabular}{c|c|c|c}
        \hline
        Variable            & Range               & Scale  & Initial value \\ \hline
        $\kappa_{x1}$       & $[0.98, 1.00]$
                            & linear              & 1.0                    \\
        $\kappa_{x2}$       & $[0.98, 1.00]$      & linear & $1.0$         \\
        $d_{x1}^{\nu}$      & $[10^{2}, 10^{10}]$ & log    & $10^6$        \\
        $d_{x2}^{\nu}$      & $[10^{2}, 10^{10}]$ & log    & $10^6$        \\
        $\alpha_{x1}^{\nu}$ & $[10^{5}, 10^{12}]$ & log    & $10^9$        \\
        $\alpha_{x2}^{\nu}$ & $[10^{5}, 10^{12}]$ & log    & $10^9$        \\ \hline
    \end{tabular}
    \label{tab:optimized_relaxation_parameters}
\end{table}

For each combination of $\alpha_0$ and $y$, we performed the optimization with multiple weight parameters ranging from $0.1$ to $0.9$ with $0.2$ increments. After obtaining the optimized relaxation parameters for each weight parameter, we evaluated their performance based on simulation-based attenuation and dispersion measurements as described in Section \ref{section:homogeneous_attenuation_measurements}. The attenuation results were assessed using the normalized root mean square error (NRMSE) between the simulated and target attenuation coefficients. The relaxation parameters that yielded the lowest NRMSE were selected as the best fit for each $\alpha_0$ and $y$ combination. In the selection, absolute dispersion error was used as a secondary criterion to ensure that the selected parameters also provided an acceptable level of accuracy in dispersion modeling.
If the NRMSE exceeded $10\%$ or absolute dispersion error exceeded $5 \text{m/s}$, we labeled the optimization as failed for that specific combination of $\alpha_0$ and $y$.

\subsection{Simulation implementation details}
The simulation framework Fullwave 2.5 is built upon the Fullwave 2~\citep{Sode2026-to} simulation platform, which has been extended to incorporate the relaxation parameters selection acquired through the optimization process described in the previous section.
The Fullwave 2.5 simulation is implemented in CUDA with Python bindings with multi-GPU support.
Fullwave 2.5 employs the finite-difference time-domain (FDTD) method to solve the stretched pressure-velocity equation (\ref{eq:wave1}) and (\ref{eq:wave2}).
In order to ensure numerical stability and accuracy in heterogeneous media with high contrast,
same as Fullwave 2~\citep{Sode2026-to}, Fullwave 2.5 utilizes the staggered-grid finite difference (FD) discretization~\citep{Tan2014-qt}, whose FD operator has 2$M$-th order accuracy in space and fourth-order accuracy in time. We empoloyed $M=4$ for the FD operator, which is an eighth-order accurate operator in space and fourth-order accurate in time.
The simulation is performed mainly on a single NVIDIA RTX 4090 GPU with 24 GB of memory. The simulation domain is discretized with 16 points per wavelength (ppw) and a Courant-Friedrichs-Lewy (CFL) number of 0.2, otherwise specified.
These parameters (16 PPW, CFL = 0.2) are used for all quantitative validation experiments in Sections \ref{subsec:homogeneous_validation}, \ref{subsec:heterogeneous_validation}, \ref{subsec:pml_boundary_stability}, and \ref{subsec:ablation_studies}  to ensure high accuracy benchmarks.
For the 3D simulation in \ref{subsec:3d_demo}, the simulation was performed with 2 NVIDIA H100 GPUs with domain decomposition in the axial direction. The simulation domain is discretized with 12 PPW and a CFL number of 0.4 to balance accuracy and computational efficiency for the large-scale 3D simulation.

\section{Validation Design}

Validation is performed by comparing simulation results with analytical solutions for attenuation and dispersion to evaluate the accuracy of the multiple relaxation model. Additionally, we performed reflection coefficient measurements and stability evaluations of the heterogeneous relaxation parameters to ensure the stability of the two-stage C-PML boundary conditions proposed in \citep{Sode2026-to}.

\subsection{Homogeneous attenuation coefficient measurements} \label{section:homogeneous_attenuation_measurements}

\begin{figure}
    \centering
    \includegraphics[width=0.9\linewidth]{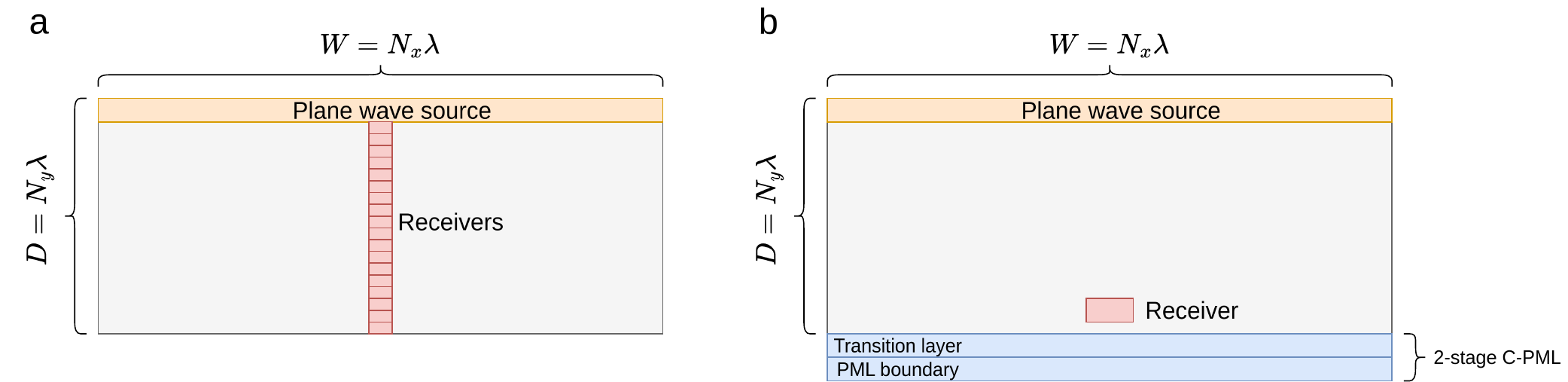}
    \caption{Simulation domain for attenuation coefficient, dispersion measurements, and reflection coefficient measurements.
        (a) Attenuation coefficient and dispersion measurement domain. A plane wave source is placed at the top of the simulation domain, and receivers are placed along the centerline of the simulation domain to measure the pressure wavefield.
        (b) Reflection coefficient measurement domain. A plane wave source is placed at the top of the simulation domain, and receivers are placed near the bottom of the simulation domain to measure the incident and reflected wave amplitudes.
        The C-PML and transition layer are applied at the bottom of the simulation domain to induce the reflection from the boundary.
    }
    \label{fig:attenuation_evaluation_domain}
\end{figure}

We used a simulation domain with a source at the top and receivers along the centerline in order to measure the attenuation coefficient for each frequency. In the simulation, we used a plane wave source that emitted waves of different frequencies. Gaussian modulated sinusoidal pulses were used as the source waveforms.

Figure \ref{fig:attenuation_evaluation_domain}a shows the simulation domain for the attenuation coefficient measurements. In the experiment, we changed the center frequency from $1$ MHz to $20$ MHz and measured the attenuation coefficient for each base attenuation coefficient $\alpha_0$ and power law exponent $y$ at each frequency. The center frequency was sampled at even intervals of $2.71$ MHz.

We employed $D=10 \lambda$ and $W=50 \lambda$ for attenuation coefficient measurements, where $\lambda$ is the wavelength of the transmitted wave at the center frequency. This domain size is chosen to minimize the edge effects and ensure accurate measurements of the attenuation coefficient. The receiver pixels are placed at distances of $0.05D$ and $0.8D$ from the source position for measuring $I_\text{reference}$ and $I_\text{attenuated}$, respectively.

We selected two receivers at a distance $d$ apart and measured the intensity at each receiver position, denoted as $I_\text{reference}$ and $I_\text{attenuated}$. The attenuation coefficient is calculated for each frequency using the following equation:
\begin{align}
    \alpha = -\cfrac{1}{2d} \ln\left(\cfrac{I_\text{attenuated}}{I_\text{reference}}\right) \quad [\text{Np}/\text{m}]
\end{align}
where $d$ is the distance between receivers.

The accuracy of the attenuation coefficient measurements is quantified using normalized root mean square error (NRMSE), defined as:
\begin{align}
    \text{NRMSE} = \frac{\text{RMSE}}{\alpha_\text{max} - \alpha_\text{min}}
\end{align}
where $\alpha_\text{max}$ and $\alpha_\text{min}$ are the maximum and minimum values of the target attenuation coefficients across all frequencies, and RMSE is the root mean square error between the simulated and target attenuation coefficients.

\subsection{Homogeneous dispersion measurements}

Similar to the attenuation coefficient measurements, we measure the dispersion by placing a source at the top of the simulation domain and receivers along the centerline of the simulation domain. The sources emit plane waves with different frequencies, and we measured the temporal pressure wave at the receiver positions. The dispersion value is calculated based on the time-of-flight of the pressure at the receiver positions.

Figure \ref{fig:attenuation_evaluation_domain}a shows the simulation domain for the dispersion measurements. In the experiment, we changed the transmission frequencies from $1$ MHz to $20$ MHz and calculated the phase velocity at each frequency.

The phase velocity is calculated as:
\begin{align}
    c = \cfrac{d}{\Delta t} \quad [\text{m}/\text{s}]
\end{align}
where $d$ is the distance between the source and receiver, and $\Delta t$ is the time-of-flight of the pressure wave at the receiver position. We employed $D=10 \lambda$ and $W=50 \lambda$ for dispersion measurements, where $\lambda$ is the wavelength of the transmitted wave at the center frequency. The receivers are placed at distances of $0.05D$ and $0.8D$ from the source position for measuring the time-of-flight.

\subsection{Reflection coefficient measurements}

Reflection coefficient is measured by placing a source at the top of the simulation domain and receivers near the bottom of the simulation domain. The source emits plane waves, and we measure the pressure wavefield at the receiver positions. The reflection coefficient is calculated based on the incident wave and the reflected wave amplitudes at the receiver positions in decibels (dB). Figure \ref{fig:attenuation_evaluation_domain}b shows the simulation domain for the reflection coefficient measurements.

Reflection coefficient is calculated as:
\begin{align}
    R = 20 \log_{10}\left(\cfrac{p_\text{reflected}}{p_\text{incident}}\right) \quad [\text{dB}] \label{eq:reflection_coefficient_pml}
\end{align}
where $p_\text{incident}$ is the amplitude of the incident wave, and $p_\text{reflected}$ is the amplitude of the reflected wave. For the simulation domain, we employed $D=5 \lambda$ and $W=20 \lambda$, where $\lambda$ is the wavelength of the transmitted wave at the center frequency. The center frequency of the plane wave source is set to $1$ MHz.

\subsection{Heterogeneous attenuation coefficient measurements with 2 layer model}

This experiment validates the Fullwave 2.5 simulation in a heterogeneous medium with two layers of different attenuation properties. The simulation domain consists of two layers with different attenuation coefficients and power law exponents. A plane wave source is placed at the top of the simulation domain, and receivers are placed along the centerline of the simulation domain to measure the pressure wavefield.

The sound speed and density are kept constant throughout the domain, while the attenuation properties vary between the two layers. The attenuation coefficient is measured using the same method described in Section \ref{section:homogeneous_attenuation_measurements}, but within each layer separately.

Physically, the dispersion difference between the two layers induces reflections at the layer interface, which can affect the accuracy of the attenuation measurements. However, the dispersion differences in biological tissues are typically small (e.g., less than 5 m/s at 1 MHz between muscle and fat). Therefore, we expect that we will observe small reflection at the layer interface in this experiment.

We have conducted three different two-layer simulations with varying attenuation properties for each layer. The attenuation properties for each layer are selected based on typical values observed in abdominal tissues, such as fat, muscle, and liver. The attenuation properties for each layer are as follows: Fat layer: $\alpha_0 = 0.4$ dB/cm/MHz$^{1.1}$, $y = 1.1$; Muscle layer: $\alpha_0 = 0.15$ dB/cm/MHz$^{1.0}$, $y = 1.0$; Liver layer: $\alpha_0 = 0.5$ dB/cm/MHz$^{1.1}$, $y = 1.1$.

\subsection{Heterogeneous relaxation parameters stability in abdominal tissue simulation}

This experiment is designed to evaluate the stability of the heterogeneous relaxation parameters in the Fullwave 2.5 simulation. In the practical application, the relaxation parameters are heterogeneous and vary across the simulation domain. We performed a simulation with heterogeneous power law distribution based on the abdominal tissue properties to demonstrate the stability of the heterogeneous relaxation parameters and the two-stage C-PML boundary conditions.

The simulation is performed in 3D. Abdominal tissue domain is constructed based on the Visible Human Project dataset \citep{Ackerman1998-xp, Spitzer1996-ly}. The tissue segmentation and preprocessing steps are identical to those described by \citet{Zhuang2025-ym}. The material properties are assigned based on the literature \citep{Lin1987-nn, Chen1987-bz, Sehgal1986-aj, Errabolu1987-du, Korta_Martiartu2021-lq, Edwards1988-pp, Youssef2018-up, Zhuang2025-ym}.

Table~\ref{tab:material_properties} summarizes the acoustic properties used for modeling muscle, fat, liver, and skin tissues in the 3D abdominal simulation.

\begin{table}[t]
    \centering
    \caption{Acoustic material properties used for modeling muscle, fat, liver, and skin tissues in the 3D abdominal simulation.}
    \label{tab:material_properties}
    \begin{tabular}{lccccc}
        \hline
        Tissue & $c$ (m/s) & $\rho$ (kg/m$^3$) & $\alpha_0$ (dB/MHz$^y$/cm) & $y$ & Reference                                      \\
        \hline
        Fat    & 1412      & 937               & 0.40                       & 1.1 & \citep{Errabolu1987-du}                        \\
        Muscle & 1527      & 1070              & 0.15                       & 1.0 & \citep{Korta_Martiartu2021-lq}                 \\
        Liver  & 1566      & 1064              & 0.50                       & 1.1 & \citep{Lin1987-nn, Chen1987-bz, Sehgal1986-aj} \\
        Skin   & 1772      & 1090              & 2.1                        & 1.0 & \citep{Edwards1988-pp, Youssef2018-up}         \\
        \hline
    \end{tabular}
\end{table}

\subsection{Adaptive versus homogeneous PML across the tissue range} \label{subsec:pml_hom_vs_het_methods}

This experiment isolates the contribution of the adaptive (per-pixel) PML tuning by comparing it against a single fixed (homogeneous) PML configuration across the full practical $(\alpha_0, y)$ range.

The simulation domain, source, and reflection-coefficient measurement protocol are identical to Section~\ref{section:homogeneous_attenuation_measurements}, and the $(\alpha_0, y)$ sweep grid is the same one used for the adaptive-PML baseline in Section~\ref{subsec:pml_boundary_stability} (Fig.~\ref{fig:ref_coeff_vs_attenuation}). The interior medium is homogeneous within each run, so no inter-layer reflection contributes to the measurement.

Three PML configurations are compared at the boundary. The first is the adaptive PML baseline, in which relaxation parameters within the PML and the transition layer are assigned per pixel from the lookup tables of Section~\ref{subsec:optimization_framework}, matching the local interior tissue $(\alpha_0, y)$. This uses the two-stage formulation of \citet{Sode2026-to} with $L_{\text{Transition}} = 3\lambda$ and $L_{\text{Bound}} = 3\lambda$, and is the configuration reported in Section~\ref{subsec:pml_boundary_stability}. The second is a homogeneous PML with transition layer, which uses the same two-stage geometry ($L_{\text{Transition}} = 3\lambda$, $L_{\text{Bound}} = 3\lambda$), but with the relaxation parameters fixed at the values optimized for $(\alpha_0, y) = (0.5, 1.0)$ throughout the absorbing region, regardless of the interior tissue. The third is a homogeneous PML without transition layer, in which only the PML region ($L_{\text{Bound}} = 3\lambda$) is present, again with parameters fixed at $(\alpha_0, y) = (0.5, 1.0)$. In this last case, the higher-order relaxation parameters that the two-stage formulation normally ramps across the transition region are instead ramped within the PML region so that the formulation remains well-defined.

For each PML configuration and each $(\alpha_0, y)$ on the sweep grid, the reflection coefficient is measured using Eq.~(\ref{eq:reflection_coefficient_pml}). We additionally monitor the peak interior pressure at every time step to detect numerical instability (NaN or Inf in the recorded wavefield), which is reported separately as a failure mode for the homogeneous configurations.

\section{Results}

\subsection{Homogeneous Validation: Attenuation and Dispersion Accuracy} \label{subsec:homogeneous_validation}

\begin{figure}
    \centering
    \includegraphics[width=1.0\linewidth]{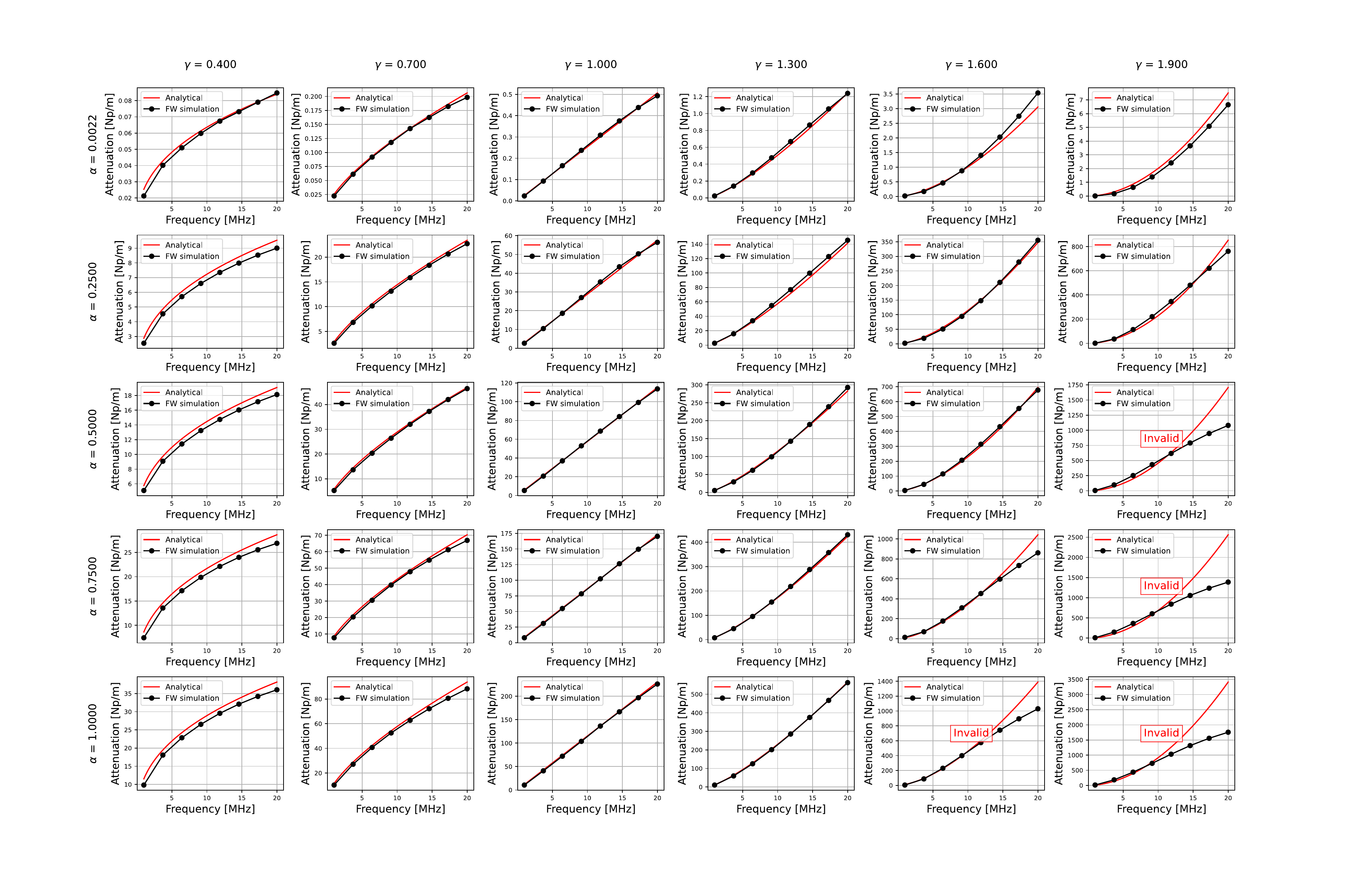}
    \caption{Comparison between the simulated and theoretical attenuation strength for different attenuation coefficients and exponents. The normalized RMSE between the simulated and theoretical attenuation coefficients was shown to be less than $5\%$ in most cases within the target frequency range of $1$--$20$ MHz. It shows the capability of the proposed method to flexibly model wide range of attenuation coefficients and exponents.}
    \label{fig:attenuation_alpha_power}
\end{figure}

\begin{figure}
    \begin{minipage}[t]{0.48\textwidth}
        \centering
        \includegraphics[width=1\linewidth]{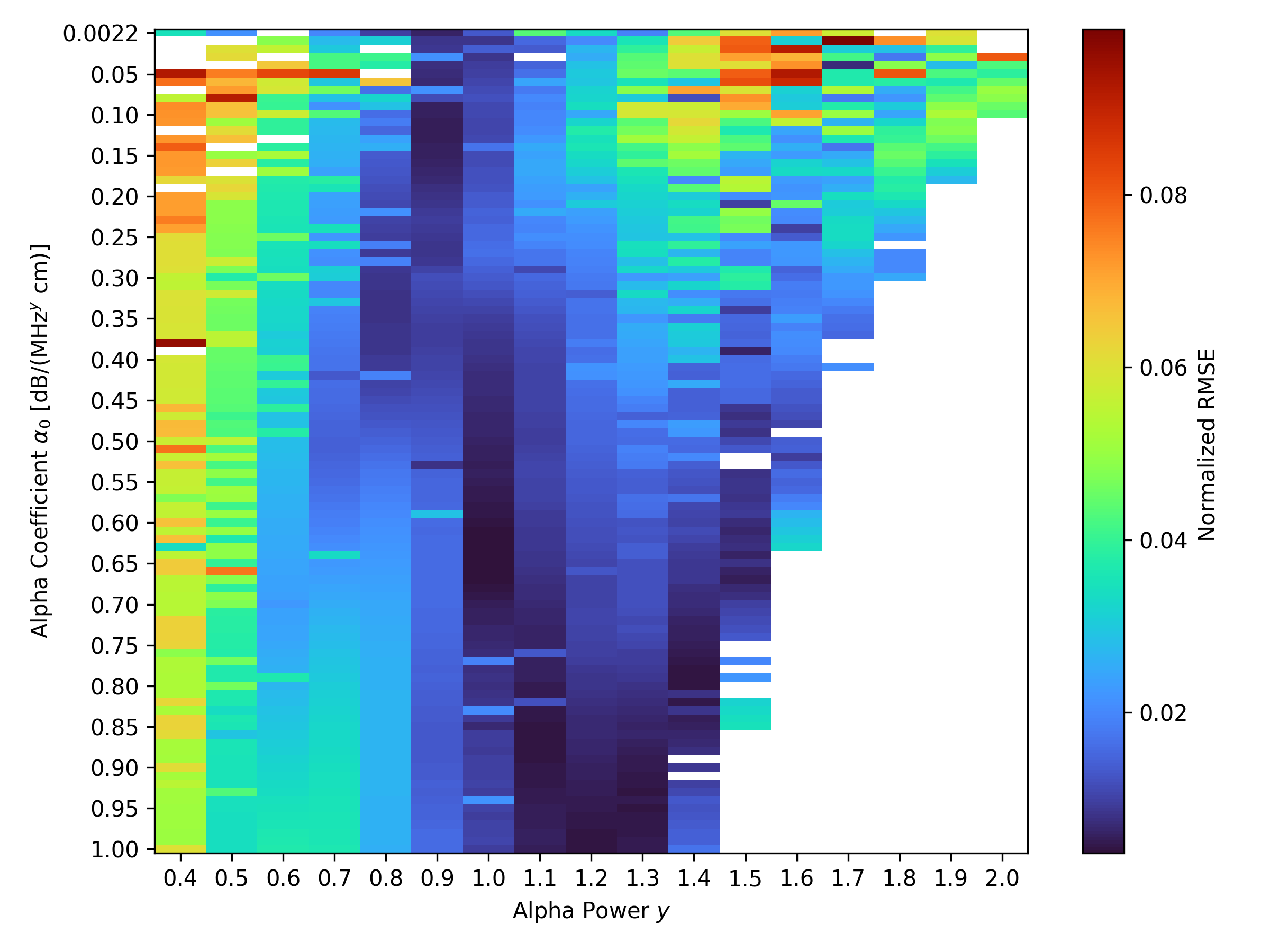}
        \subcaption{normalized RMSE heatmap for all $\alpha_0$ and $y$ cases. Invalid cases with $\text{NRMSE} > 10\%$ are shown in white.}
        \label{fig:error_matrix_heatmap}
    \end{minipage}
    \begin{minipage}[t]{0.48\textwidth}
        \centering
        \includegraphics[width=1\linewidth]{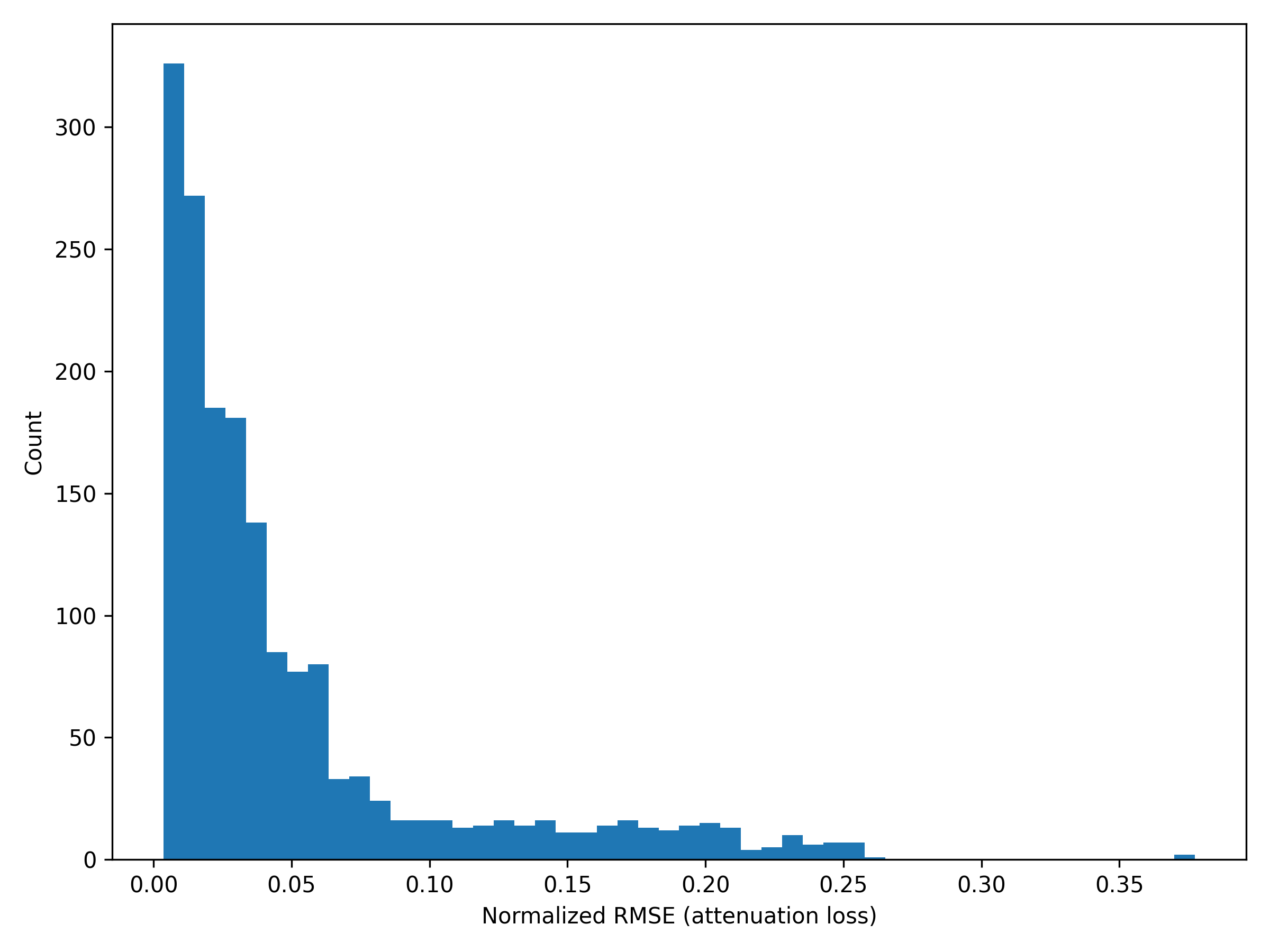}
        \subcaption{histogram of normalized RMSE for all cases}
        \label{fig:error_matrix_histogram}
    \end{minipage}
    \caption{(a) Normalized RMSE between the simulated and theoretical attenuation coefficients across the full optimization range ($\alpha_0 = 0.0022$--$1.0$ dB/(MHz$^y$ cm), $y = 0.4$--$2.0$, 1328 combinations). Invalid cases with $\text{NRMSE} > 10\%$ are shown in white. The core region ($y = 0.7$--$1.4$) achieves NRMSE $< 10\%$ for the full $\alpha_0$ range. (b) Histogram of normalized RMSE for all cases. The figure shows that most cases achieved an NRMSE of less than $5\%$, demonstrating the effectiveness of the proposed optimization method across a wide range of attenuation parameters.}

    \label{fig:error_matrix}

\end{figure}

The optimization results for $\alpha=0.0022, 0.25, 0.50, 0.75$ and $y=0.40, 0.70, 1.0, 1.3, 1.6$ are shown in Figure \ref{fig:attenuation_alpha_power}.
Two relaxation parameters ($N=2$ in Eq \ref{eq:relaxation4}) were used for the optimization.
The target frequency range was set to $1-20 \ \text{MHz}$, which is a typical frequency range for medical ultrasound applications.
The figure shows the comparison between the simulated and theoretical attenuation strength for different attenuation coefficients and exponents.
The case with $\alpha=0.5$ and $y=1.0$ is a typical attenuation model for human tissue.
Figure \ref{fig:error_matrix_heatmap} shows the normalized RMSE between the simulated and theoretical attenuation coefficients for all $\alpha_0$ and $y$ cases spanning the full optimization range ($\alpha_0 = 0.0022$--$1.0$ dB/(MHz$^y$ cm), $y = 0.4$--$2.0$). Invalid cases with $\text{NRMSE} > 10\%$ are shown in white.
The clinically relevant core region ($\alpha_0 = 0.0022$--$1.0$, $y = 0.7$--$1.4$) mostly achieves NRMSE $< 10\%$ across the full attenuation coefficient range.
For higher exponents ($y \geq 1.5$), the validated attenuation coefficient range progressively narrows: $y=1.5$ supports $\alpha_0 < 0.75$, $y=1.6$ supports $\alpha_0 < 0.62$, and $y=1.7$ supports $\alpha_0 < 0.35$.
The pairs with higher coefficients and higher exponents tend to show higher NRMSE due to the increased complexity of fitting steep frequency-dependent attenuation curves.
Figure \ref{fig:error_matrix_histogram} shows the histogram of normalized RMSE for all cases.
While some cases show higher NRMSE, the histogram shows that most cases achieved an NRMSE of less than $5\%$, demonstrating the effectiveness of the proposed optimization method across a wide range of attenuation parameters.
The mean and standard deviation of the NRMSE for all valid cases are $2.61\%$ and $1.83\%$, respectively.
Note that the cases in the range $y=0.7$--$1.4$ show excellent accuracy with NRMSE typically less than $2\%$. These attenuation exponents are commonly observed in biological soft tissues and are particularly relevant for medical ultrasound applications.

\begin{figure}
    \centering
    \includegraphics[width=1.0\linewidth]{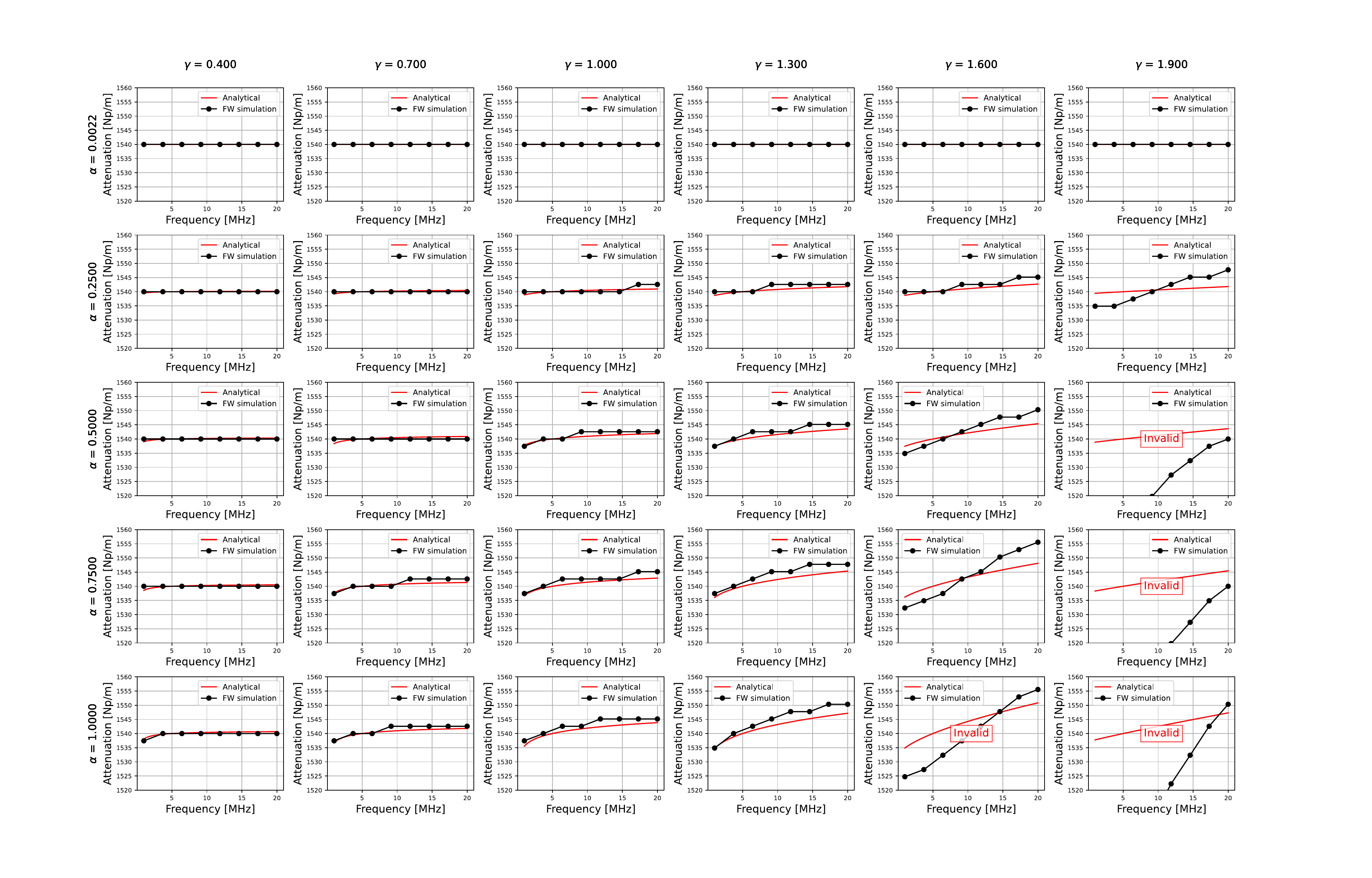}
    \caption{Comparison between the simulated and theoretical dispersion for different attenuation coefficients and exponents. The dispersion varies in average of $1.14 \pm 0.76 \text{m/s}$, which is $\pm 0.07\%$ of the sound speed value of $1540 \text{m/s}$.}
    \label{fig:dispersion_alpha_power}
\end{figure}

Figure \ref{fig:dispersion_alpha_power} shows the comparison between the simulated and theoretical dispersion for different attenuation coefficients and exponents. While the dispersion was used as a regularization term in the optimization, the figure shows that the proposed method can also accurately model dispersion.
The RMSE of dispersion was $1.14 \pm 0.76 \text{m/s}$, which is $\pm 0.07\%$ of the sound speed value of $1540 \text{m/s}$.

\subsection{Heterogeneous Validation: Two-Layer Models} \label{subsec:heterogeneous_validation}

\begin{figure}
    \centering
    \begin{minipage}[t]{0.32\textwidth}
        \centering
        \includegraphics[width=1.0\linewidth]{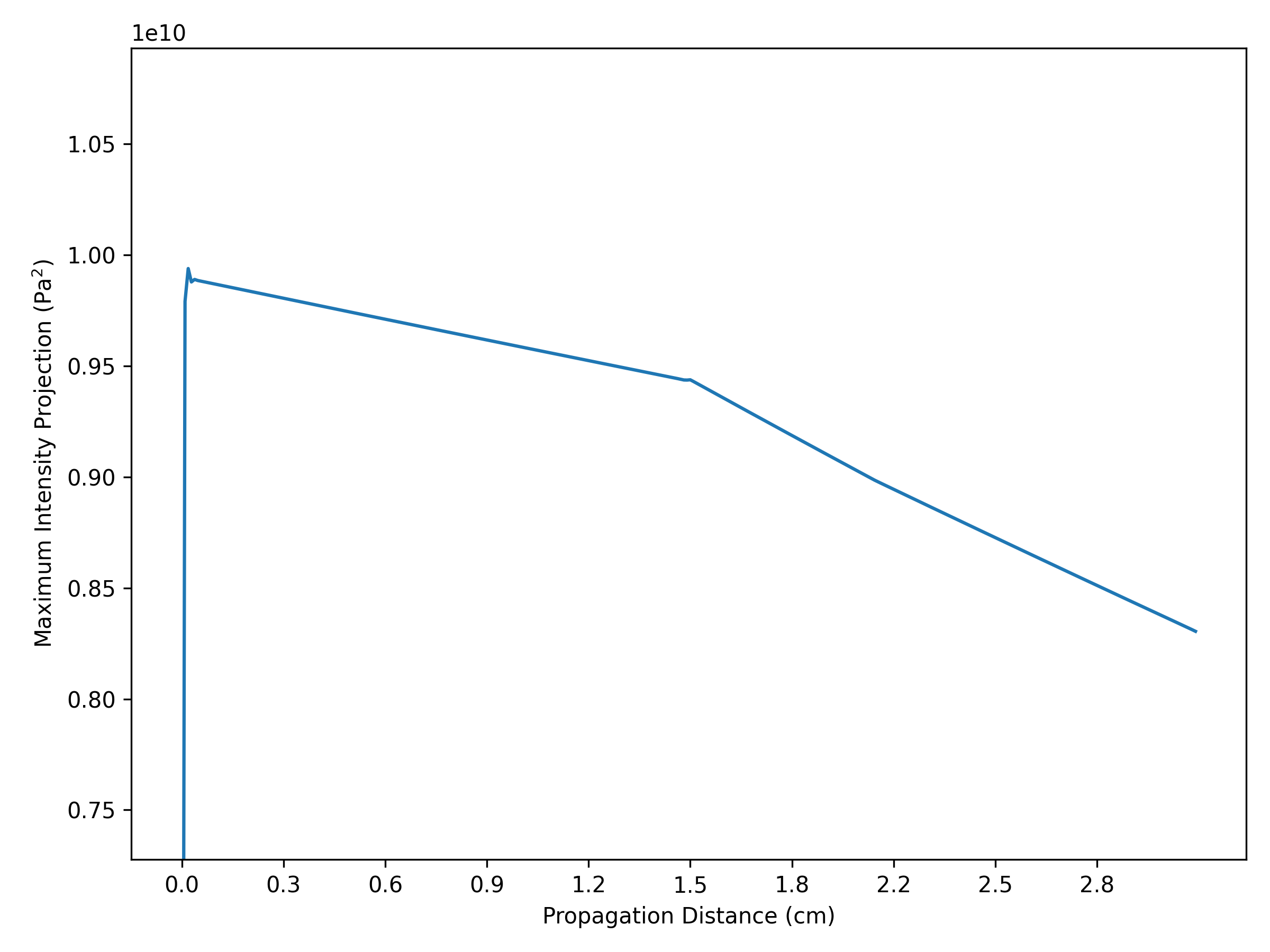}
        \subcaption{muscle to fat:\\$\alpha=0.15, y=1.0$ to $\alpha=0.40, y=1.1$}
        \label{fig:heterogeneous_validation_results_a}
    \end{minipage}\hfill
    \begin{minipage}[t]{0.32\textwidth}
        \centering
        \includegraphics[width=1.0\linewidth]{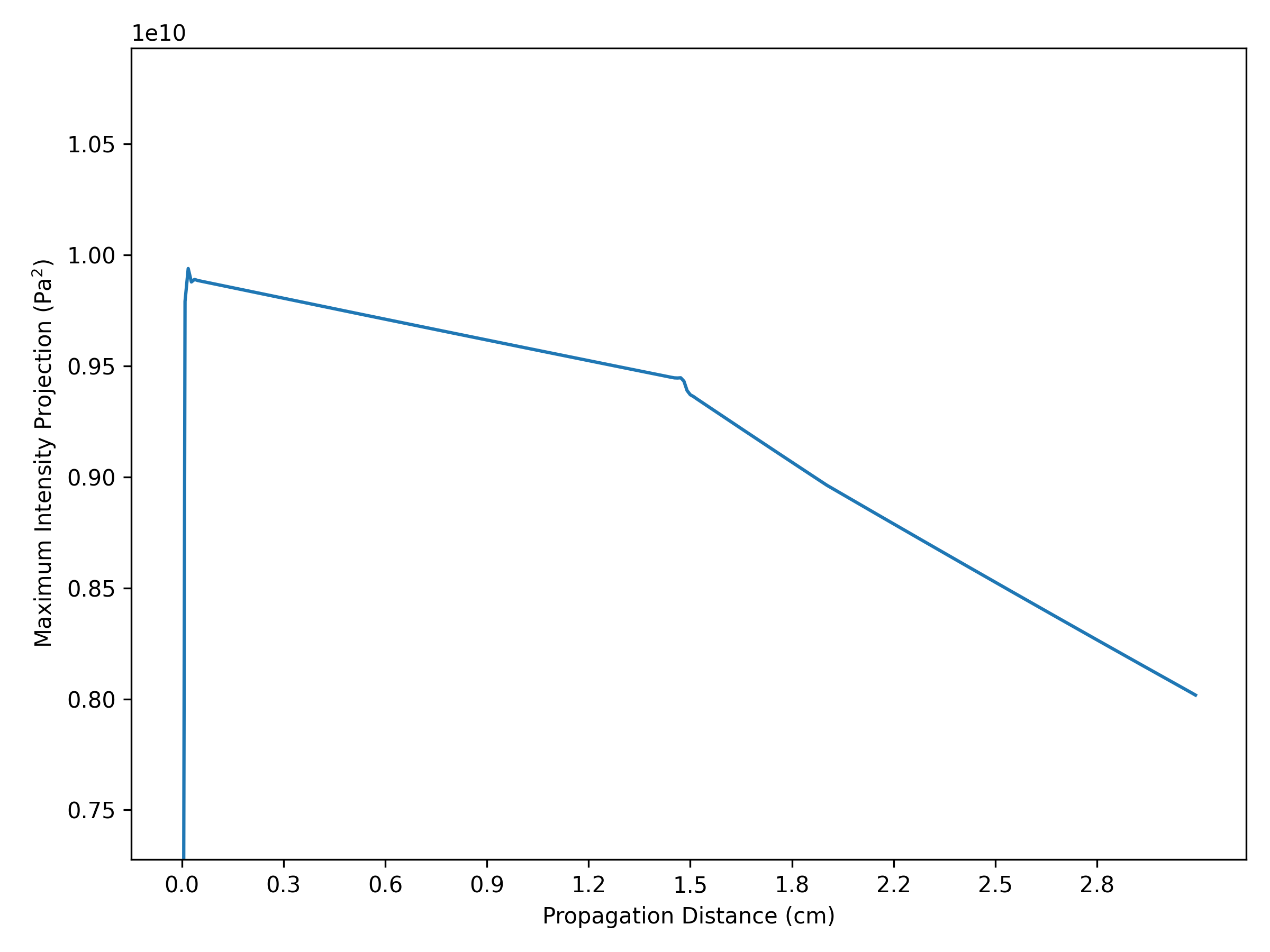}
        \subcaption{muscle to liver:\\$\alpha=0.15, y=1.0$ to $\alpha=0.50, y=1.1$}
        \label{fig:heterogeneous_validation_results_b}
    \end{minipage}\hfill
    \begin{minipage}[t]{0.32\textwidth}
        \centering
        \includegraphics[width=1.0\linewidth]{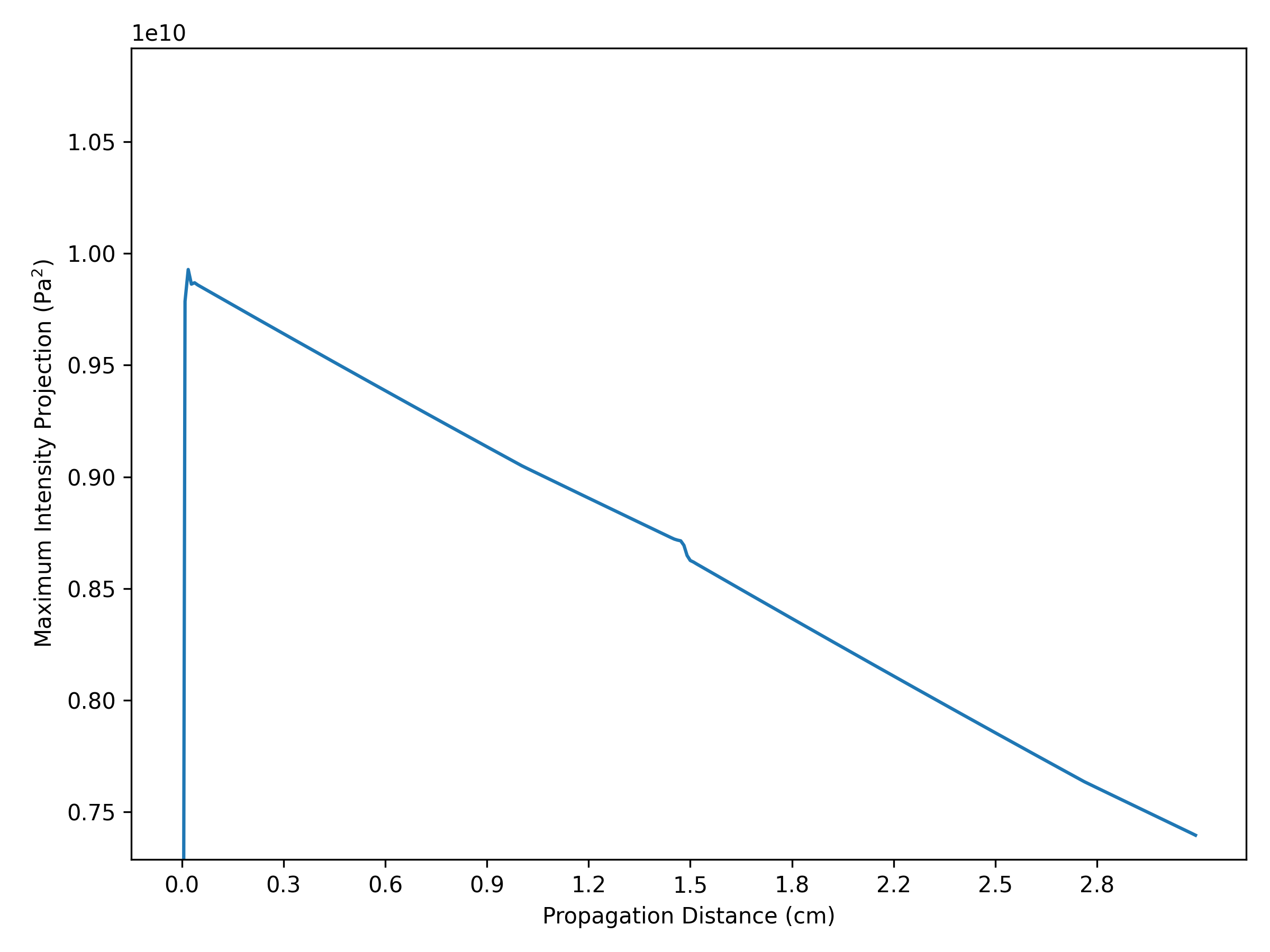}
        \subcaption{fat to liver:\\$\alpha=0.40, y=1.1$ to $\alpha=0.50, y=1.1$}
        \label{fig:heterogeneous_validation_results_c}
    \end{minipage}\hfill

    \caption{Maximum intensity projection (MIP) at the center axis of the simulation domain when the wave propagates through a two-layer heterogeneous medium with different attenuation parameters corresponding to muscle, fat, and liver tissue properties. (a) muscle to fat: $\alpha=0.15, y=1.0$ to $\alpha=0.40, y=1.1$. (b) muscle to liver: $\alpha=0.15, y=1.0$ to $\alpha=0.50, y=1.1$. (c) fat to liver: $\alpha=0.40, y=1.1$ to $\alpha=0.50, y=1.1$. The sound speed and density were kept homogeneous to isolate dispersion effects. The figure shows that the wave propagates stably before and after the interface. The small reflection observed at the interface is due to the dispersion mismatch caused by the different power-law attenuation parameters.}

    \label{fig:heterogeneous_validation_results}
\end{figure}

Figure \ref{fig:heterogeneous_validation_results} shows the maximum intensity projection (MIP) at the center axis of the simulation domain when the wave propagates through a two-layer heterogeneous medium with different attenuation parameters. We can see that the wave propagates stably before and after the interface of different attenuation parameters, and the amplitude decay changes according to the attenuation parameters of each layer.

We measured the reflection coefficient at the interface of different attenuation parameters to evaluate the physical accuracy of the proposed method in heterogeneous media.
Three different combinations of attenuation parameters corresponding to muscle, fat, and liver tissue properties were tested: (a) muscle to fat: $\alpha=0.15, y=1.0$ to $\alpha=0.40, y=1.1$, (b) muscle to liver: $\alpha=0.15, y=1.0$ to $\alpha=0.50, y=1.1$, and (c) fat to liver: $\alpha=0.40, y=1.1$ to $\alpha=0.50, y=1.1$. To isolate the effects of dispersion, the sound speed and density were kept homogeneous across the interface, so any observed reflections are purely due to the dispersion mismatch caused by different power-law attenuation parameters. Each simulation was performed with a $5 \ \text{MHz}$ Gaussian modulated sinusoidal pulse as the input.
For (a), the measured reflection coefficient at the interface was $-63.0 \ \text{dB}$, for (b) it was $-50.2 \ \text{dB}$, and for (c) it was $-50.5 \ \text{dB}$. Since the acoustic impedance ($Z = \rho c$) is identical on both sides of the interface, the theoretical impedance-based reflection would be zero. The observed small reflections are entirely due to the dispersion-induced phase velocity mismatch between the two media with different power-law exponents and coefficients. These results demonstrate that the C-PML formulation correctly captures the dispersive nature of the medium while maintaining numerical stability at heterogeneous interfaces.

The attenuation coefficient was measured separately in each layer using the method described in Section \ref{section:homogeneous_attenuation_measurements}. Table~\ref{tab:two_layer_validation} summarizes the quantitative validation results for all three tissue combinations. The NRMSE values for both layers remain below 2.5\%, and the interface reflection coefficients are below $-50$ dB. These results confirm that the method accurately preserves attenuation properties across heterogeneous interfaces.

\begin{table}[t]
    \centering
    \caption{Quantitative validation results for two-layer heterogeneous media. The NRMSE values represent the normalized error in attenuation coefficient measurements for each layer. Interface reflection coefficients are measured at 5 MHz.}
    \begin{tabular}{lccc}
        \hline
        Layer configuration        & Layer 1 NRMSE & Layer 2 NRMSE & Interface reflection (dB) \\
        \hline
        Muscle $\rightarrow$ Fat   & 1.18\%        & 0.847\%       & $-63.0$                   \\
        Muscle $\rightarrow$ Liver & 1.18\%        & 1.41\%        & $-50.2$                   \\
        Fat $\rightarrow$ Liver    & 0.978\%       & 2.32\%        & $-50.5$                   \\
        \hline
    \end{tabular}
    \label{tab:two_layer_validation}
\end{table}

\subsection{3D Realistic Tissue Simulation: Stability and Scalability Validation} \label{subsec:3d_demo}
\begin{figure}
    \centering
    \begin{minipage}[t]{0.48\textwidth}
        \centering
        \includegraphics[width=\linewidth]{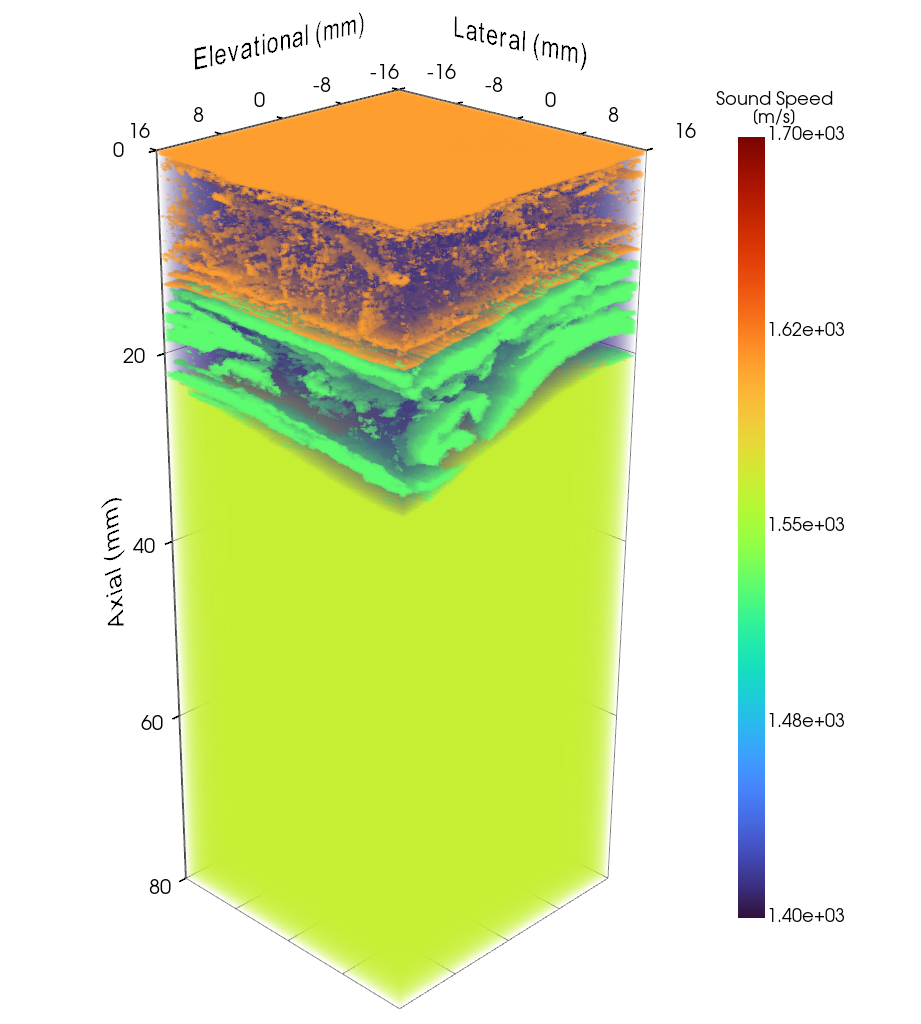}
        \subcaption{Three-dimensional sound speed $c$ map.}
        \label{fig:heterogeneous_simulation_sound_speed}
    \end{minipage}
    \begin{minipage}[t]{0.48\textwidth}
        \centering
        \includegraphics[width=\linewidth]{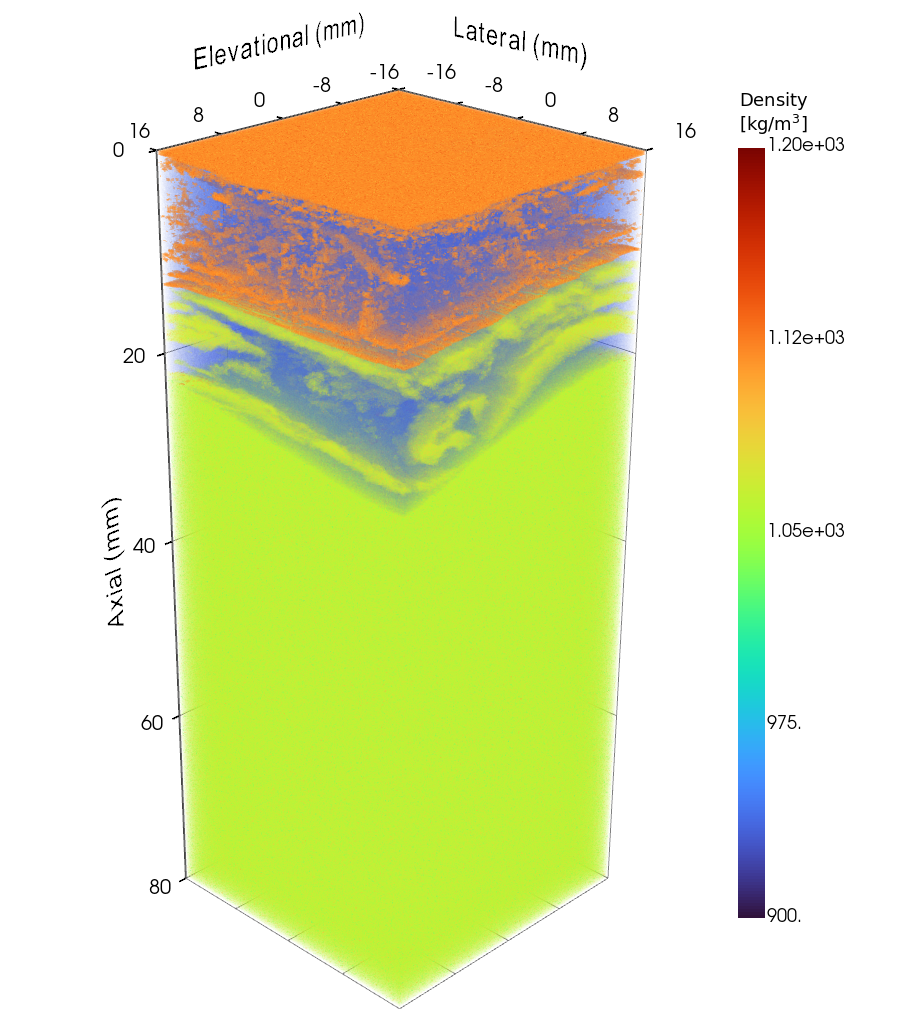}
        \subcaption{Three-dimensional density $\rho$ map.}
        \label{fig:heterogeneous_simulation_density}
    \end{minipage}
    \begin{minipage}[t]{0.48\textwidth}
        \centering
        \includegraphics[width=\linewidth]{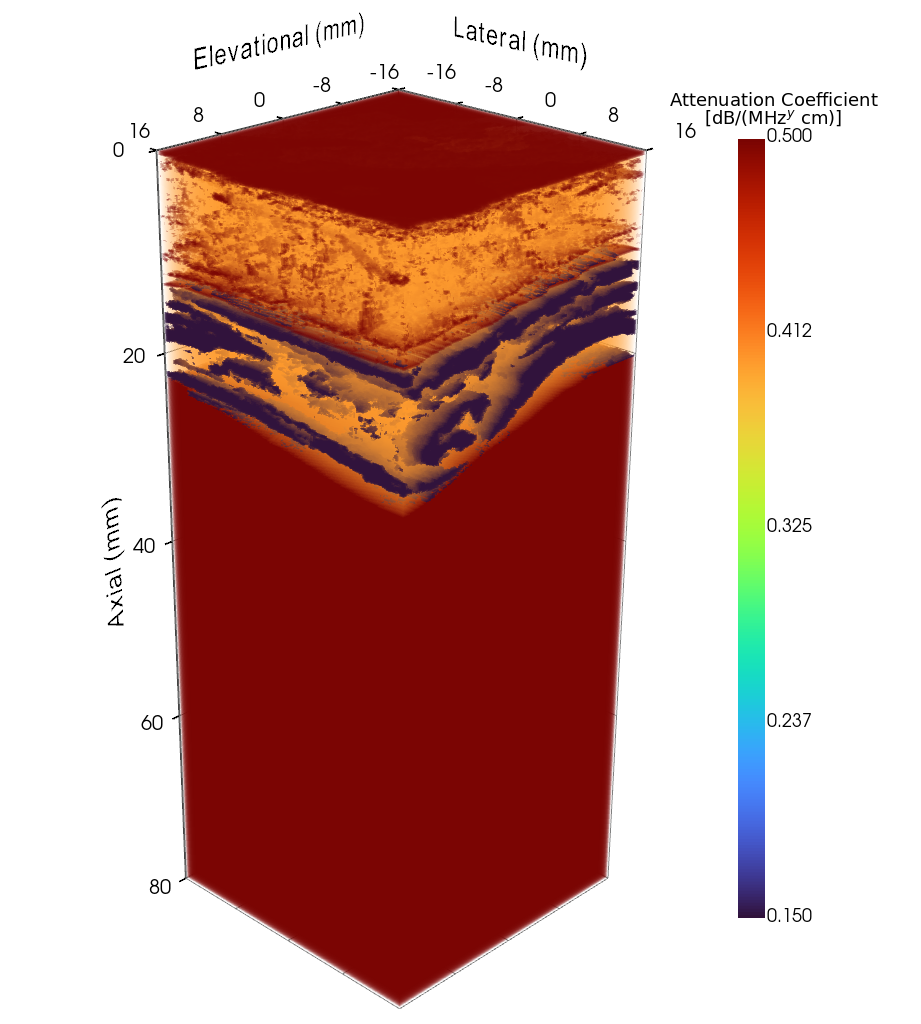}
        \subcaption{Three-dimensional attenuation coefficient $\alpha_0$ map.}
        \label{fig:heterogeneous_simulation_attenuation_coefficient}
    \end{minipage}
    \begin{minipage}[t]{0.48\textwidth}
        \centering
        \includegraphics[width=\linewidth]{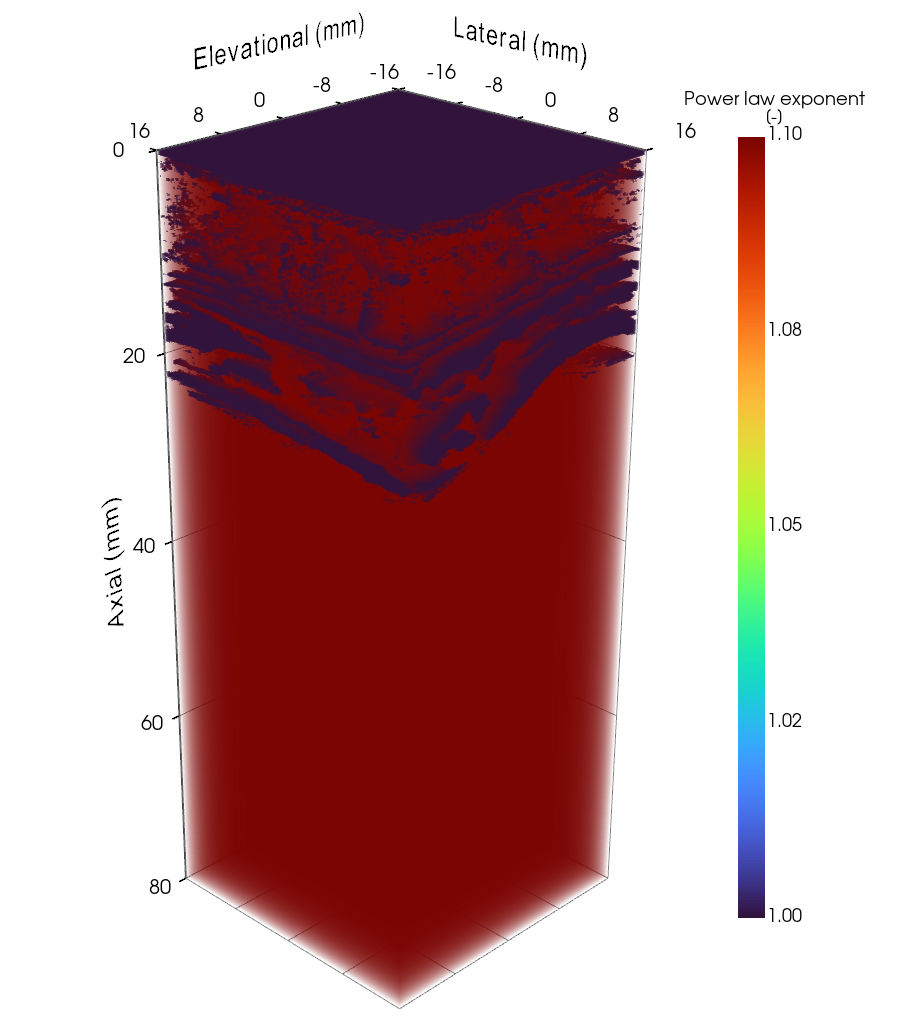}
        \subcaption{Three-dimensional attenuation exponent $y$ map.}
        \label{fig:heterogeneous_simulation_attenuation_exponent}
    \end{minipage}
    \caption{The input heterogeneous medium to demonstrate the simulation of heterogeneous power-law attenuation. It models an abdominal wall and liver.}
    \label{fig:heterogeneous_simulation_domain}

\end{figure}

\begin{figure}
    \centering
    \begin{minipage}[t]{0.48\textwidth}
        \centering
        \includegraphics[width=\linewidth]{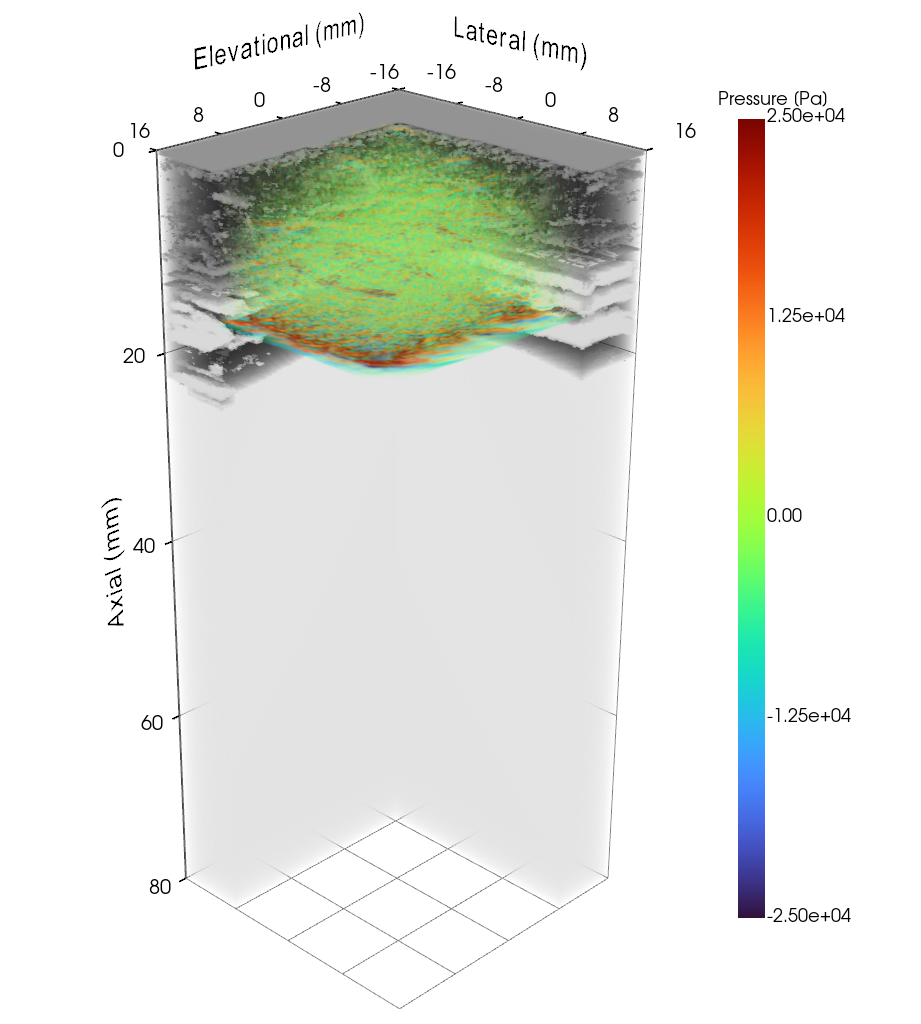}
        \subcaption{$t=2.5$ \textmu s
        }
        \label{fig:abdominal_simulation_wave_propagation_0}
    \end{minipage}
    \begin{minipage}[t]{0.48\textwidth}
        \centering
        \includegraphics[width=\linewidth]{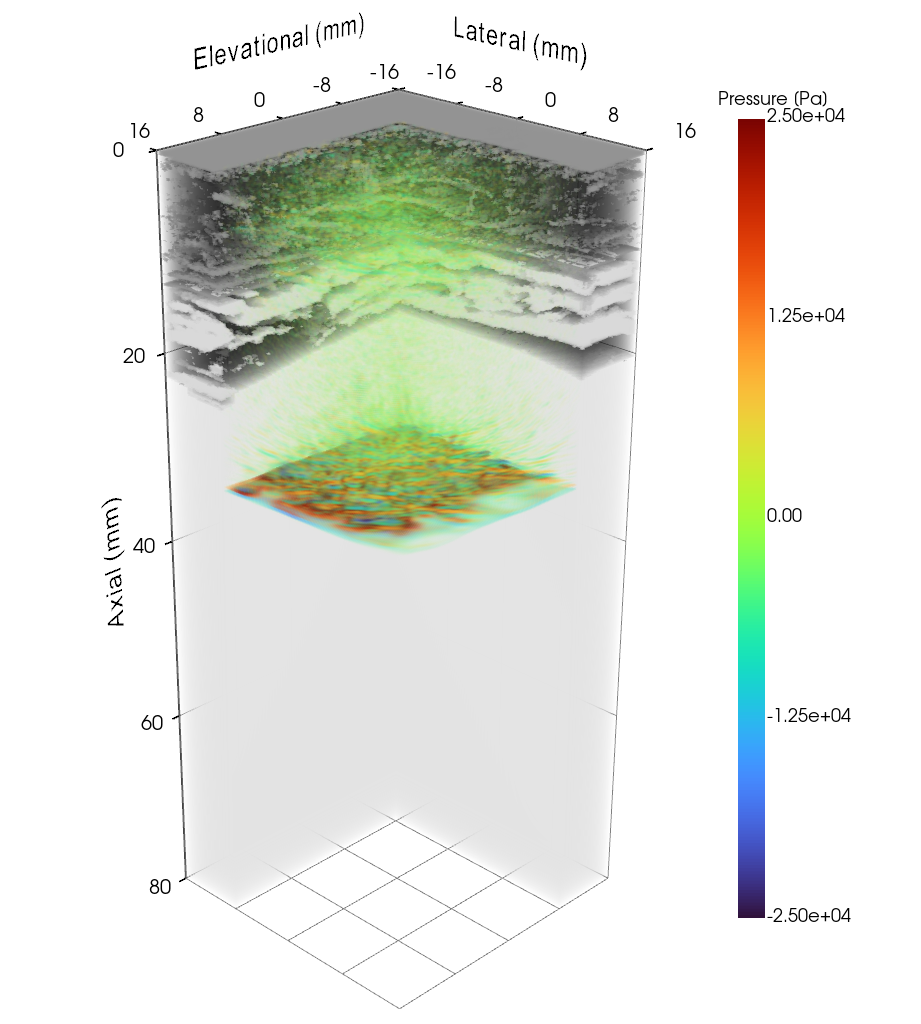}
        \subcaption{$t=5.0$ \textmu s
        }
        \label{fig:abdominal_simulation_wave_propagation_1}
    \end{minipage}\hfill
    \caption{The wave propagation in the abdominal wall and liver model with heterogeneous power-law attenuation. The figure shows that the wave propagates stably without any instability or artifacts.}
    \label{fig:abdominal_simulation_wave_propagation}
\end{figure}%

\begin{figure}
    \centering
    \begin{minipage}[t]{0.43\textwidth}
        \centering
        \includegraphics[width=\linewidth]{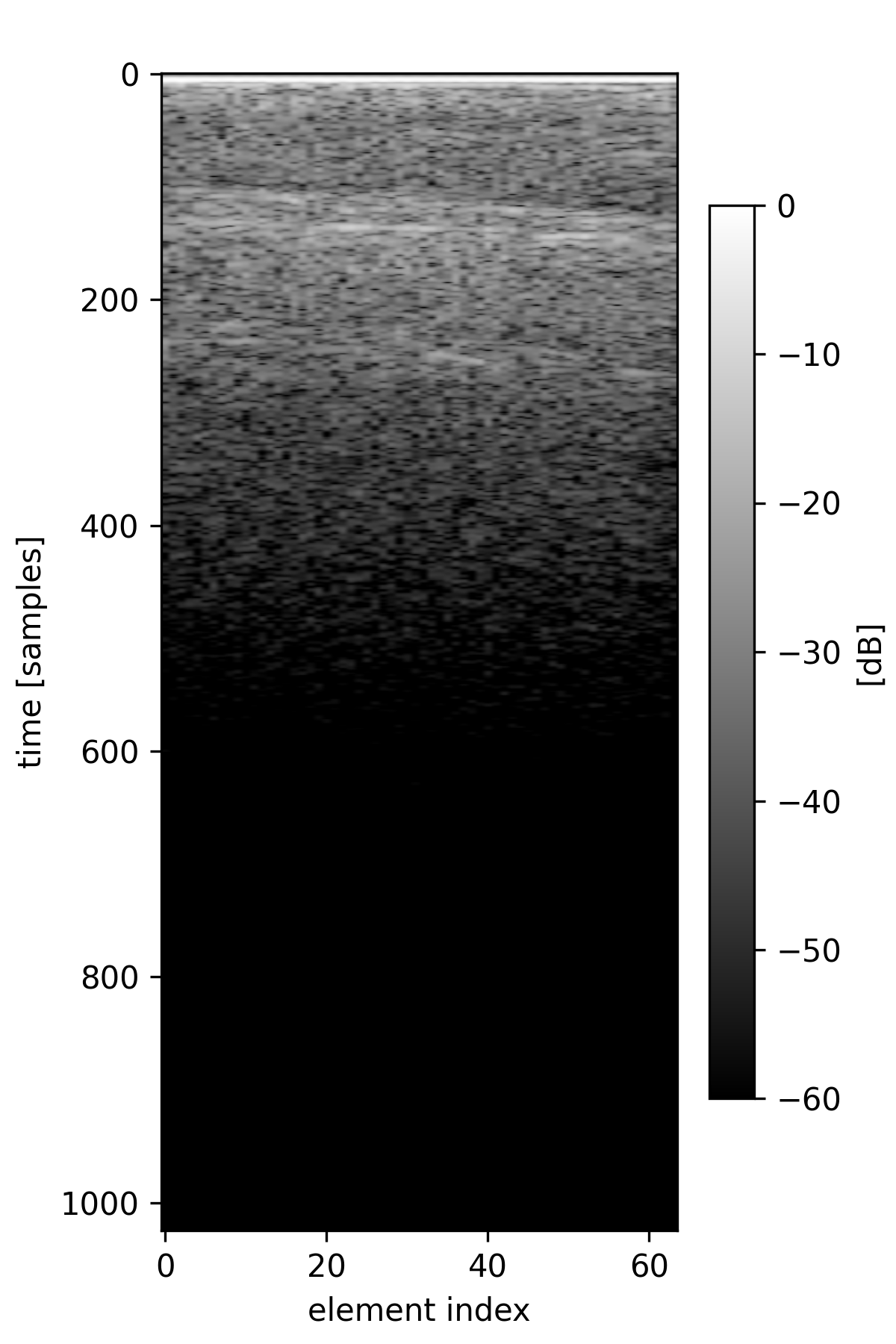}
        \subcaption{RF data}
        \label{fig:abdominal_simulation_rf_image}
    \end{minipage} \hfill
    \begin{minipage}[t]{0.43\textwidth}
        \centering
        \includegraphics[width=\linewidth]{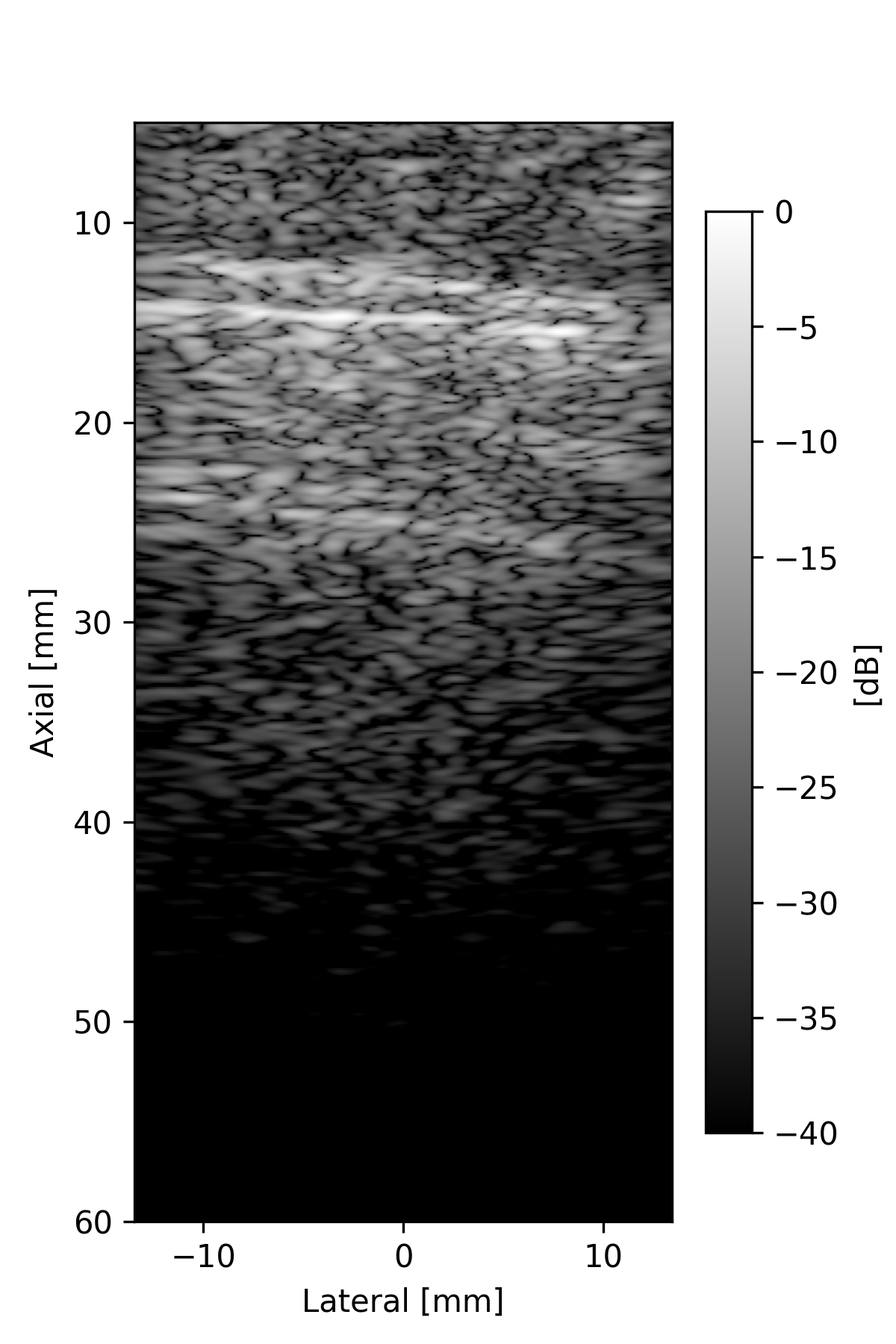}
        \subcaption{Reconstructed B-mode image}
        \label{fig:abdominal_simulation_bmode_image}
    \end{minipage}
    \caption{RF data and reconstructed B-mode image obtained from the 3D simulation with heterogeneous power-law attenuation.}
    \label{fig:abdominal_simulation_rf_and_bmode}
\end{figure}%

As a practical application of the proposed method, we demonstrate the simulation of heterogeneous power-law attenuation.
We prepared a 3D abdominal wall and liver model (Figure \ref{fig:heterogeneous_simulation_domain}) as a simulation target and modeled a P4-1C transducer to run the simulation.
We used a segmented Visible Human Project (VHP) dataset~\citep{Ackerman1998-xp,Zhuang2025-ym} for the abdominal wall and liver model. The liver used the material values stated in \ref{tab:material_properties}. The simulation domain size was $32 \ \text{mm} \times 32 \ \text{mm} \times 80 \ \text{mm}$, and the transducer was placed at the center of the top surface of the simulation domain. The transducer was excited with a $2 \ \text{MHz}$ Gaussian modulated sinusoidal pulse. The transducer aperture size was  $24 \ \text{mm} \times 20 \ \text{mm}$, and the plane wave sequence was transmitted at an angle of $0^\circ$.
The 3D abdominal simulation uses 12 PPW and CFL = 0.4 to satisfy GPU memory constraints with two H100 GPUs; this discretization is within the stability regime confirmed by the CFL condition. The simulation was performed with a time step of  $16.7 \ \text{ns}$ and a total simulation time of $0.119 \ \text{ms}$. The simulation grid size was $499 \times 499 \times 1247$.
The large grid size (499 $\times$ 499 $\times$ 1247) required 110 GB of GPU memory. The simulation was executed using 2 NVIDIA H100 GPUs with domain decomposition, with a total execution time of approximately 9 minutes.

Figure \ref{fig:abdominal_simulation_wave_propagation} shows the wave propagation in the abdominal wall and liver model with heterogeneous power-law attenuation. The figure shows that the wave propagates stably without any instability, even when subresolutional heterogeneous power-law attenuation is assigned.

Figure \ref{fig:abdominal_simulation_rf_and_bmode} shows the RF data and reconstructed B-mode image obtained from the simulation. The figure shows that the RF data and B-mode image are successfully obtained with heterogeneous power-law attenuation assigned.

We can observe the multiple reverberations caused by the abdominal wall below 25 mm depth in the B-mode image. This phenomenon is commonly observed in abdominal ultrasound imaging due to the strong impedance mismatch between the transducer, skin, and subcutaneous fat layers. The simulation successfully reproduces this effect, indicating that the proposed method can accurately model complex wave interactions in heterogeneous media with varying power-law attenuation.

The two-layer quantitative validation (Table~\ref{tab:two_layer_validation}, $<2.5\%$ NRMSE per layer) establishes accuracy for well-defined planar interfaces where analytical solutions exist. The 3D abdominal simulation demonstrates three critical capabilities: (1) scalability to realistic complex anatomical geometries, (2) numerical stability with voxel-level variation in both $\alpha_0(\mathbf{x})$ and $y(\mathbf{x})$, and (3) the first time-domain simulation with spatially heterogeneous power-law exponents $y(\mathbf{x})$. The observed reverberations in the B-mode image (Fig. \ref{fig:abdominal_simulation_rf_and_bmode}) demonstrate that the simulation correctly captures complex wave interactions including multiple scattering and aberration through heterogeneous layers. See Section 6.2 for validation strategy rationale.

\subsection{PML Boundary Stability} \label{subsec:pml_boundary_stability}

We evaluated the reflection coefficient of the two-stage PML proposed in~\citep{Sode2026-to} when varying the attenuation parameters across different $\alpha_0$ and $y$ combinations. The C-PML and transition layer thickness were set to $3 \lambda$ and $3 \lambda$, respectively, where $\lambda$ is the wavelength.
\begin{figure}[htbp]
    \centering
    \includegraphics[width=0.8\linewidth]{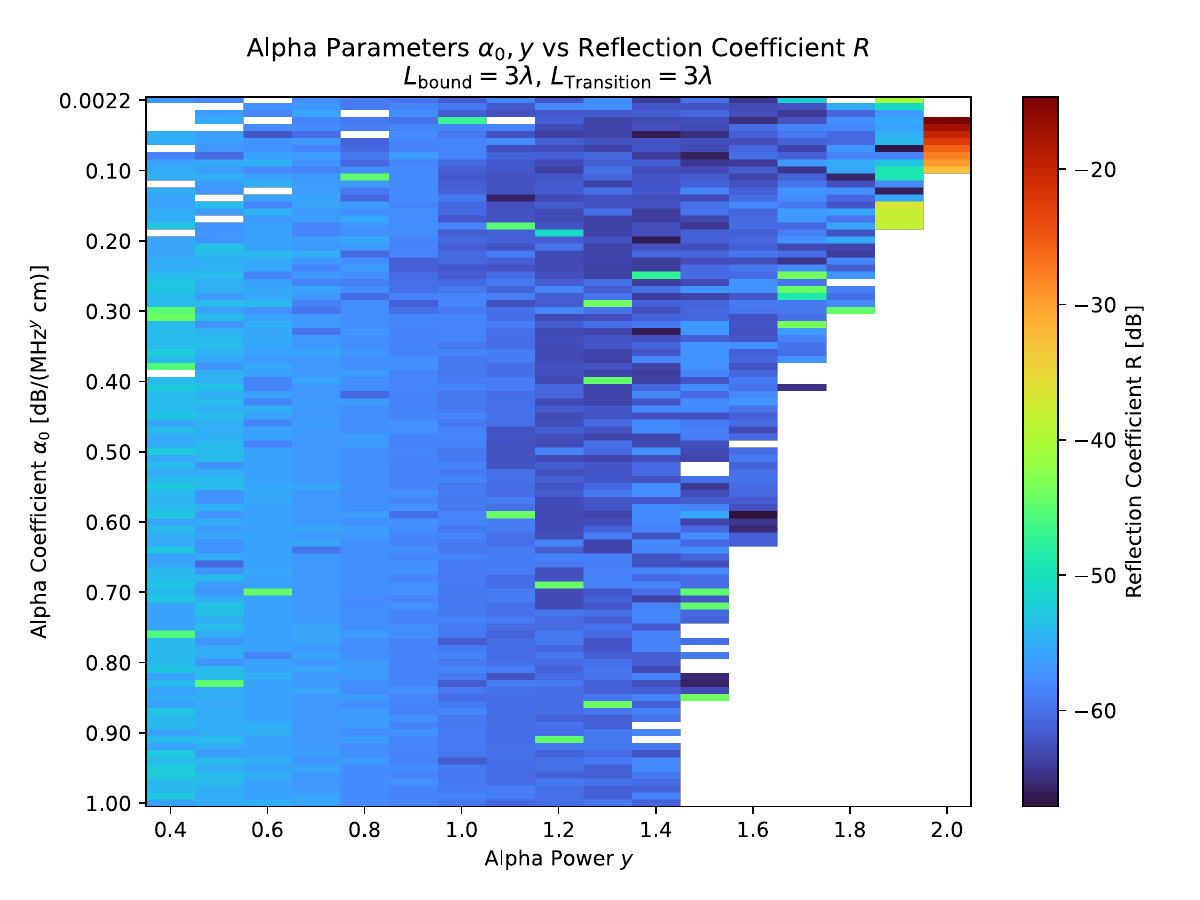}
    \caption{Reflection coefficient experiments for the two stage PML with transition layer when varying attenuation parameters. The PML and transition layer thickness are set to $L_{\text{Transition}} = 3\lambda$ and $L_{\text{Bound}} = 3\lambda$. In most cases, the reflection coefficient is less than -50 dB, demonstrating the stability of the two-stage PML against a broad range of attenuation parameters used in medical ultrasound simulations.}
    \label{fig:ref_coeff_vs_attenuation}
\end{figure}
Figure \ref{fig:ref_coeff_vs_attenuation} shows the reflection coefficient when the attenuation coefficient and attenuation exponent are varied.
The relaxation parameter sets failed to be optimized were excluded from the plot.
The reflection coefficient tend to increase as the attenuation exponent increases, while it is relatively stable against changes in the attenuation coefficient.
Especially for high attenuation exponents such as $y \geq 1.9$, the reflection coefficient increases.
For the higher attenuation exponents, such as $y = 2.0$, the reflection coefficient exceeds -20 dB, which may not be acceptable for some ultrasound simulations.
We need to further improve the PML formulation to reduce the reflection coefficient for high attenuation exponents in future work.
However, for most medical ultrasound applications, the attenuation exponent is less than 1.8, and the reflection coefficient is less than -50 dB, which is acceptable for most ultrasound simulations.

This heatmap also demonstrates the stability of the proposed two-stage PML against the broad range of attenuation parameters used in medical ultrasound simulations.

\subsection{Adaptive versus homogeneous PML across the tissue range} \label{subsec:pml_hom_vs_het}

To isolate the contribution of the adaptive PML tuning, we repeated the reflection-coefficient sweep of Section~\ref{subsec:pml_boundary_stability} with two homogeneous PML variants, following the design described in Section~\ref{subsec:pml_hom_vs_het_methods}: a fixed PML at $(\alpha_0, y) = (0.5, 1.0)$ with the same two-stage geometry as the adaptive baseline, and the same fixed PML without the transition layer.

\begin{figure}[t]
    \centering
    \includegraphics[width=0.6\linewidth]{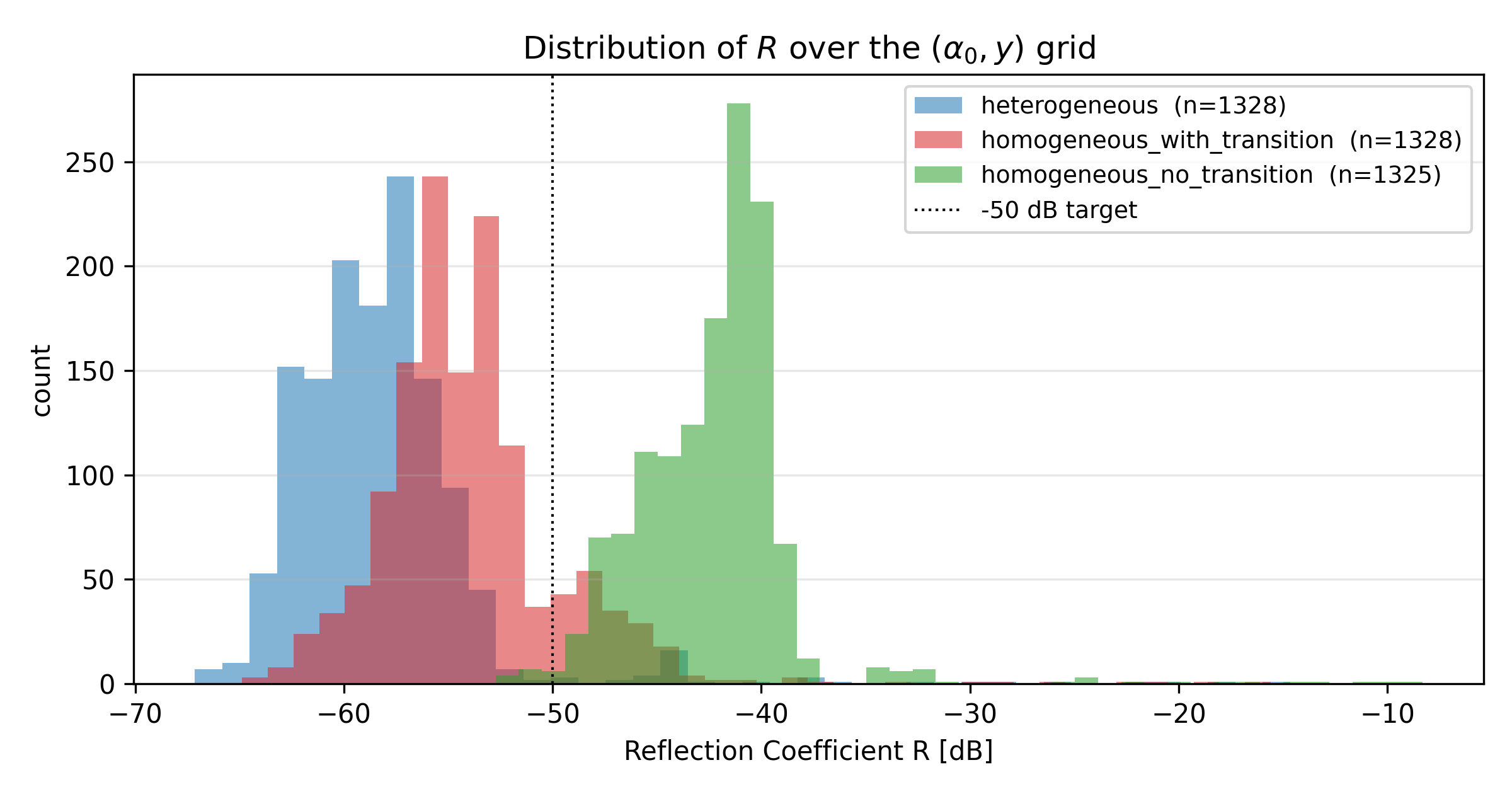}
    \caption{Distribution of the reflection coefficient $R$ over the $(\alpha_0, y)$ sweep grid for three PML configurations: the adaptive PML used throughout the rest of this work (blue), a homogeneous PML fixed at $(\alpha_0, y) = (0.5, 1.0)$ with the same $3\lambda$ transition layer (red), and the same homogeneous PML without a transition layer (green). The dotted line marks the $-50$ dB target. The adaptive PML and the homogeneous PML with transition layer both peak below the target, but the homogeneous distribution is shifted by ${\sim}4$ dB and has a tail crossing $-50$ dB. The no-transition-layer distribution is shifted by ${\sim}16$ dB and lies almost entirely above the target. The per-cell spatial structure of $R$ over $(\alpha_0, y)$ is provided in Fig.~\ref{fig:pml_hom_vs_het_heatmap} in the appendix.}
    \label{fig:pml_hom_vs_het}
\end{figure}

Figure~\ref{fig:pml_hom_vs_het} compares the distribution of measured reflection coefficients over the $(\alpha_0, y)$ sweep grid for the three configurations; the corresponding per-cell heatmaps showing the spatial structure of the degradation across the $(\alpha_0, y)$ grid are shown in Fig.~\ref{fig:pml_hom_vs_het_heatmap} in the appendix. Across the $1328$ valid $(\alpha_0, y)$ cells, the adaptive PML maintained $R < -50$ dB in $97.1\%$ of cells (median $R = -58.5$ dB), reproducing the result of Fig.~\ref{fig:ref_coeff_vs_attenuation}. The homogeneous PML with transition layer was approximately $4$ dB worse on the median ($-54.5$ dB) and met the $-50$ dB target in only $85.1\%$ of cells, with the degradation concentrated at high-$y$ values away from the design point $(0.5, 1.0)$. The homogeneous PML without transition layer was approximately $16$ dB worse than the adaptive baseline on the median ($-41.9$ dB), met the $-50$ dB target in only $1.1\%$ of cells, and produced numerical instability (NaN or Inf in the recorded wavefield) in $29$ cells distributed across the moderate-$y$ regime.

The degradation of the homogeneous configurations away from the design point reflects the dispersion mismatch between the fixed PML tuning and the interior medium. Because the multiple-relaxation formulation carries the dispersion signature of $(\alpha_0, y)$ through the spatial derivative, a PML tuned to a single $(\alpha_0, y)$ absorbs waves correctly only for that specific tissue; for any other interior tissue the dispersion-mismatched PML generates a reflected component back into the simulation domain. The adaptive PML eliminates this mismatch by inheriting the local relaxation parameters from the interior at every voxel within the absorbing region, and the transition layer further reduces the residual reflection and stabilizes the absorber against long-time energy accumulation.

Both ingredients of the two-stage adaptive PML are therefore necessary: per-voxel matching of relaxation parameters closes the median $4$ dB gap between the adaptive configuration and the homogeneous-with-transition variant, which raises the $R < -50$ dB success rate from $85.1\%$ to $97.1\%$ across the clinical range; and the transition layer is essential both for reflection performance (a further ${\sim}12$ dB on the median) and for long-time stability (eliminating the $29$ divergent cells observed without it). Removing either ingredient leaves the absorber unfit for general-purpose use across the soft-tissue parameter range.

\subsection{Ablation Studies} \label{subsec:ablation_studies}

\subsubsection{Optimization Algorithm Comparison}
\label{subsec:optimization_comparison}

\begin{figure}[t]
    \centering
    \begin{minipage}[t]{0.45\textwidth}
        \includegraphics[width=1\linewidth]{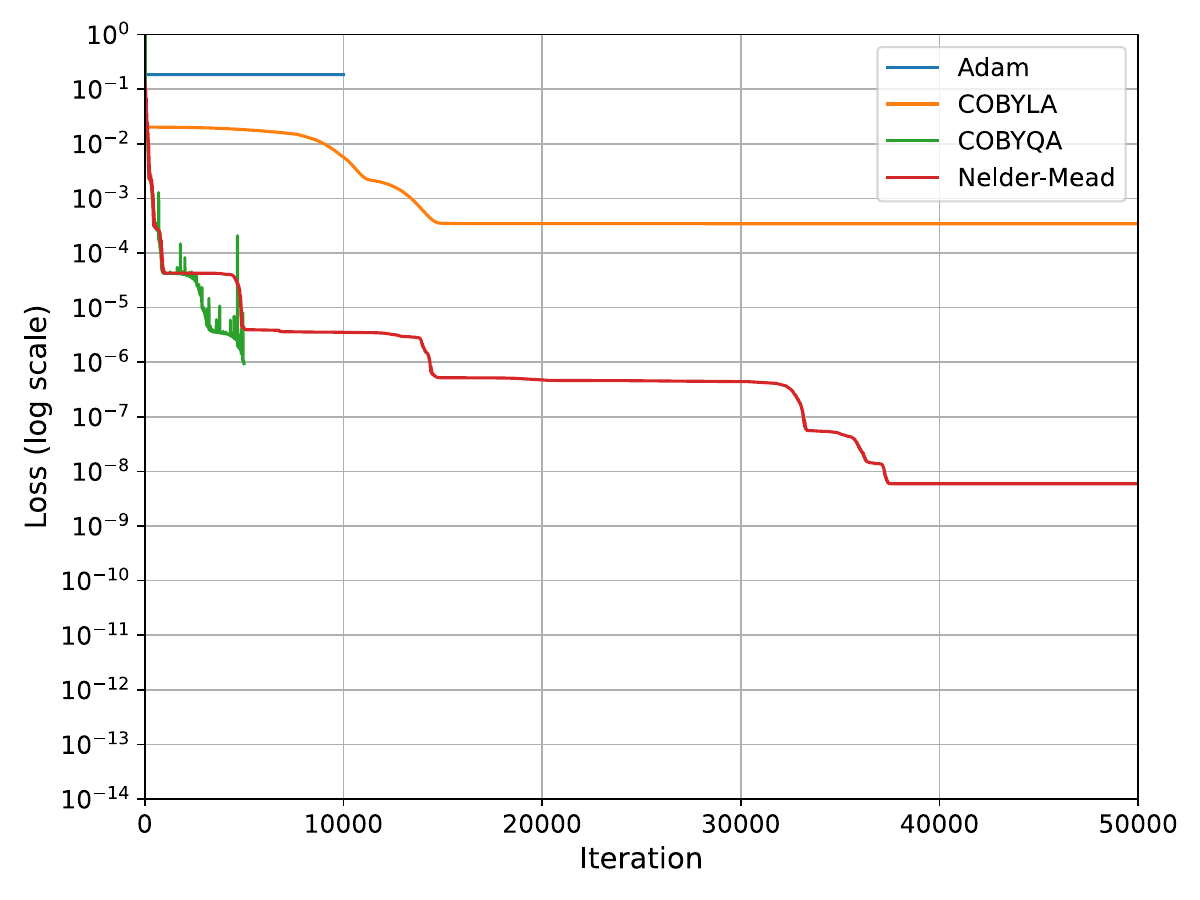}
        \subcaption{attenuation loss convergence plot for $\alpha=0.50, y=1.0, w=0.3$}
        \label{fig:optimization_algorithms_a}
    \end{minipage}\hfill
    \begin{minipage}[t]{0.45\textwidth}
        \includegraphics[width=1\linewidth]{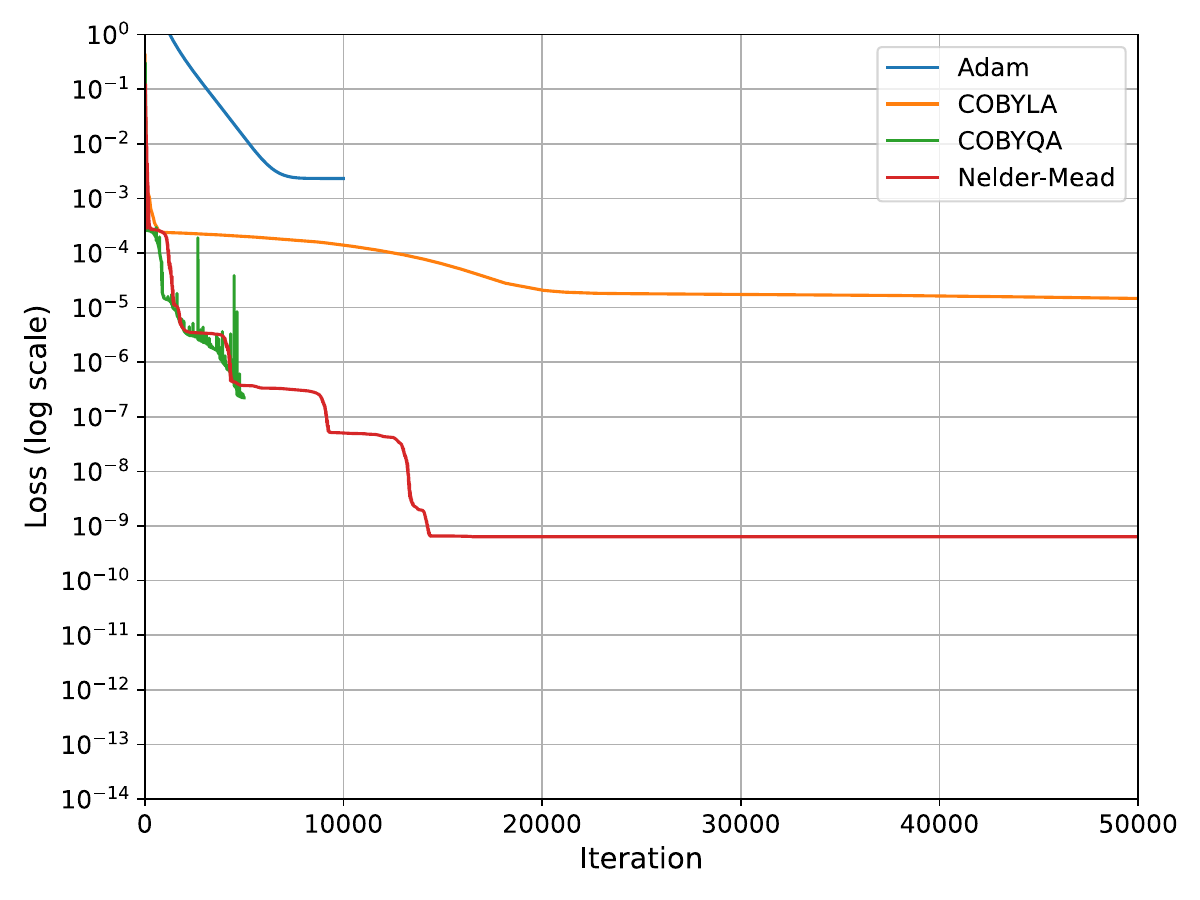}
        \subcaption{attenuation loss convergence plot for $\alpha=0.75, y=1.5, w=0.7$}
        \label{fig:optimization_algorithms_b}
    \end{minipage}\hfill

    \caption{Comparison of optimization algorithms for relaxation parameters optimization. The optimization was performed for a target power-law attenuation with $\alpha=0.50, y=1.0$ (a) and $\alpha=0.75, y=1.5$ (b).
        The figure shows the objective function value versus the number of iterations. The Adaptive Nelder-Mead algorithm converges to a lower objective function value than the other algorithms constantly.}
    \label{fig:optimization_algorithms}
\end{figure}

Figures \ref{fig:optimization_algorithms} show the optimization results for different optimization algorithms.
The optimizations were performed for a target power-law attenuation with $\alpha=0.50, y=1.0$ (Figure \ref{fig:optimization_algorithms_a}) and $\alpha=0.75, y=1.5$ (Figure \ref{fig:optimization_algorithms_b}).

The figures plot the objective function value against the number of iterations.
The loss axis is shown in a logarithmic scale to clearly show the convergence behavior of each algorithm.
As figures show, the adaptive Nelder-Mead algorithm converged to the lowest objective function value of all the algorithms.
While the other algorithms such as the COBYQA showed also good convergence behavior, they converged to a higher objective function value than the Nelder-Mead algorithm.

Nelder-Mead consistently outperformed the other algorithms in all the optimization experiments conducted in this study. The margin of improvement was typically 2-3 orders of magnitude in terms of the objective function value.
Therefore, we adopted the adaptive Nelder-Mead algorithm for this study.

\subsubsection{Effect of Number of Relaxation Parameters $N$}
\begin{figure}[t]
    \centering
    \includegraphics[width=0.9\linewidth]{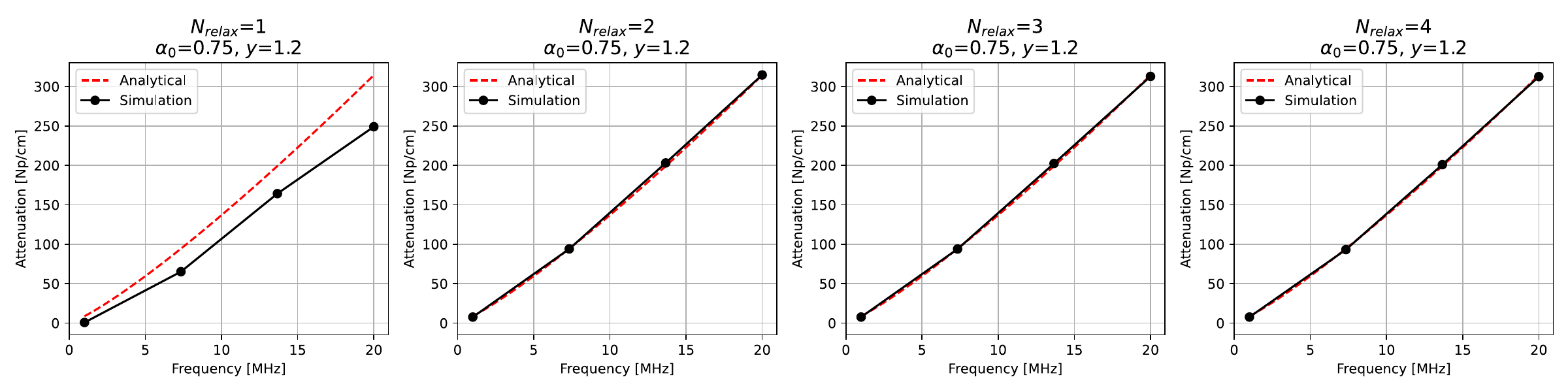}
    \caption{
        Effect of the number of relaxation parameters on attenuation modeling.
        The figure shows the comparison between the simulated and theoretical attenuation strength for different numbers of relaxation parameters for a target power-law attenuation with $\alpha=0.75, y=1.2$.
        Increasing the number of relaxation parameters from 1 to 2 improves the modeling performance in the high-frequency region. In contrast, no improvement was observed by increasing from 2 to 3 or 4.
        Thus, we use two relaxation parameters in subsequent experiments, considering the trade-off between computational cost and modeling accuracy.
    }
    \label{fig:relaxation_number_comparison}
\end{figure}

Figure \ref{fig:relaxation_number_comparison} demonstrates how the number of relaxation parameters affects attenuation modeling for a target power-law attenuation with $\alpha=0.75, y=1.2$.
Increasing the number of relaxation parameters from 1 to 2 improves the modeling performance in the high-frequency region. In contrast, no improvement was observed by increasing from 2 to 3 or 4.
Thus, we use two relaxation parameters in subsequent experiments, considering the trade-off between computational cost and modeling accuracy.
\section{Discussion}
\subsection{Optimization Strategy \& Performance}

The method accurately models a wide range of attenuation and dispersion in biological tissues by optimizing multiple relaxation parameters while maintaining numerical stability.
The optimization flexibly accommodates different attenuation coefficients and exponents.
Technically, the method fits arbitrary attenuation models, not limited to power-law. Furthermore, assigning spatial distributions of relaxation parameters enables heterogeneous and subresolution attenuation simulation, which conventional methods cannot achieve.

The optimization controls both attenuation coefficient and dispersion modeling through relaxation parameters.
The complex wavenumber-based optimization accurately models both attenuation and dispersion simultaneously.

The method achieves sufficient accuracy for most ultrasound simulations when considering measurement error in biological tissues.
For example, the reported attenuation coefficient at $1$ MHz for normal liver tissue varies from $0.47$ to $0.58$ dB/(MHz$^{1.0}$ cm)~\citep{Parker1988-ql, Lee2023-sw}, representing $\sim$20\% variability in published measurements. Given this inherent measurement uncertainty, the method's ability to achieve less than 3\% relative error provides sufficient accuracy for practical ultrasound simulation applications.

Finding a robust optimization algorithm across the full practical $(\alpha_0, y)$ range was a central engineering challenge of this work. We evaluated several derivative-free and gradient-based algorithms, including COBYQA, COBYLA, Powell, and the gradient-based Adam, in addition to the adaptive Nelder-Mead variant. Most produced attenuation curves that deviated from the analytical power-law target over substantial portions of the parameter space, particularly in the high-$\alpha_0$ and high-$y$ region where the branch-cut character of $(i\omega)^y$ (Section~\ref{subsec:theoretical_framing}) forces tight pole placement over a wide band. The adaptive Nelder-Mead variant was the only algorithm we tested that delivered acceptable attenuation fits across all 1328 tissue parameter combinations. Researchers extending the framework to new tissue regimes, broader frequency bands, or larger numbers of mechanisms should expect to revisit the algorithm choice and verify it directly on the attenuation curve.

Not every parameter combination could be fit. The 1328 cases reported above are those that passed the $10\%$ NRMSE threshold. The cases that failed appear as white cells in the high-$\alpha_0$ and high-$y$ corner of Fig.~\ref{fig:error_matrix_heatmap}. These failures are not only a limitation of our optimization procedure. They also reflect a basic property of relaxation models. As described in Section~\ref{subsec:theoretical_framing}, a true power-law attenuation $\alpha_0 f^y$ corresponds to an infinite sum of relaxation mechanisms. With a finite number $N$ of mechanisms, the model can match the power law accurately only over a limited frequency band and a limited range of $(\alpha_0, y)$. Some failures at the corners of the parameter space are therefore expected for any finite $N$.

Even within the range that can be fit, the calibration was sensitive to the choice of optimization algorithm and to the initial parameter values. A single set of initial values did not work equally well for every $(\alpha_0, y)$ pair, because the shape of the objective function changes across the parameter range. We used $N=2$ relaxation mechanisms as a deliberate compromise. This number covers most soft tissues seen in diagnostic ultrasound while keeping the optimization manageable. Using more mechanisms, or choosing the initial values from the tissue properties rather than from a fixed starting point, may extend the validated range and is a natural direction for future work.

\subsection{Validation Strategy}

The validation employs analytical solutions as the gold standard for accuracy assessment, providing more rigorous verification than numerical-to-numerical comparisons between simulators.
The two-layer validation (Table~\ref{tab:two_layer_validation}) establishes accuracy against analytical solutions for planar interfaces, while the 3D abdominal simulation demonstrates scalability to realistic anatomical geometries where multiple scattering and reverberation preclude simple analytical predictions.

\subsection{Heterogeneous Modeling Capability}

The method demonstrates numerical stability and accurate wave propagation in simulations with heterogeneous relaxation parameters.
The method enables time-domain ultrasound simulation with spatially varying power-law exponents $y(\mathbf{x})$. Prior time-domain methods (e.g., k-Wave~\citep{Treeby2010-gb}) support heterogeneous attenuation coefficients $\alpha_0(\mathbf{x})$ with a spatially uniform exponent $y$~\citep{Treeby2014-mk, Spa2025-kd}, but the capability to model spatially varying exponents $y(\mathbf{x})$ is novel to this work.
Wave propagation is accurately modeled without numerical instability, and attenuation and dispersion effects are well represented. This enables application to realistic scenarios with spatially varying tissue properties, including regions where both the attenuation coefficient and exponent vary at subwavelength scales.
The local nature of the multiple relaxation approach facilitates implementation of heterogeneous media with spatially varying attenuation properties and enables multi-GPU domain decomposition for large-scale simulations using parallel computing architectures.
This locality avoids the global FFT operations required by fractional derivative methods, which present computational challenges for domain decomposition in parallel computing environments.
This capability is important for applications such as quantitative ultrasound imaging (QUS), high-intensity focused ultrasound (HIFU) therapy, and transcranial ultrasound imaging, where accurate modeling of wave propagation in heterogeneous media is crucial for treatment planning and image reconstruction.

\subsection{PML Boundary Stability}

The two-stage C-PML formulation effectively minimizes boundary reflections while maintaining numerical stability in simulations with power-law attenuation and dispersion. For clinically relevant tissue exponents ($y \leq 1.5$), boundary reflections remain below $-50$ dB (Section~\ref{subsec:optimization_framework}). However, for high exponents approaching $y = 2.0$ (dispersionless, thermoviscous fluids), reflection coefficients increase significantly. The current C-PML was designed for dispersive media and may not be optimal for such cases.

The comparison in Section~\ref{subsec:pml_hom_vs_het} further isolates the contribution of the adaptive PML tuning. A homogeneous PML fixed at a single $(\alpha_0, y)$ matches the adaptive baseline only at the design point and degrades elsewhere on the sweep grid; the transition layer reduces but does not eliminate this degradation. The local multiple-relaxation formulation therefore relies on adaptive tuning in the PML region itself, not only in the interior, to maintain low reflection across the soft-tissue range. For simulations of heterogeneous media, where the interior tissue varies in space and the PML interface generally faces several different tissue types along its length, this argument extends voxel-by-voxel: each PML voxel needs to be tuned to its locally adjacent interior.

\subsection{Limitations \& Future Directions}

The method has several limitations that suggest directions for future work.
First, the effects of aberration and reverberation in heterogeneous media have not been quantitatively assessed using experimental phantom data. While the method has been validated against analytical solutions and demonstrates consistent behavior in realistic tissue geometries, validation with tissue-mimicking phantoms remains necessary to confirm accuracy for complex wave interactions in real-world scenarios.

Second, like Fullwave 2, the multiple relaxation approach requires more computational memory than conventional FDTD due to auxiliary memory variables for relaxation mechanisms. Memory requirements are identical to Fullwave 2 (same implementation, same number of relaxation parameters). This limits feasible simulation size and spatial resolution on a single GPU, though the multi-GPU implementation with domain decomposition mitigates constraints for moderately large simulations. Multi-node parallelization could further expand the scale of simulations, enabling full 3D modeling of large anatomical regions with high resolution.

Third, the optimization of relaxation parameters yields non-unique solutions, as different parameter sets perform better under specific conditions. Systematic investigation of the parameter space would identify robust configurations for various applications.

The method provides a versatile framework for simulating wave propagation in media with spatially varying power-law attenuation and dispersion, with applications including quantitative ultrasound imaging, therapeutic ultrasound planning, and transcranial imaging. It facilitates large-scale generation of synthetic datasets for deep learning-based ultrasound image reconstruction and can support research on inversion techniques and aberration and reverberation correction methods. %
\section{Conclusion}

We have presented Fullwave 2.5, an automated calibration framework for time-domain ultrasound simulation of spatially heterogeneous power-law attenuation. The method replaces manual grid search with derivative-free optimization, achieving mean errors below $3\%$ across diverse tissue parameter combinations spanning $\alpha_0 = 0.0022$--$1.0$ dB/(MHz$^y$ cm) and $y = 0.4$--$2.0$, with the clinically relevant core region ($y = 0.7$--$1.4$) achieving this accuracy across the full attenuation coefficient range. The method enables voxel-level variation in both attenuation coefficient $\alpha_0(\mathbf{x})$ and exponent $y(\mathbf{x})$ in time-domain FDTD. Validation through two-layer quantitative models and a 3D abdominal simulation confirms numerical stability and accuracy at clinically relevant tissue interfaces.

This capability unlocks applications previously inaccessible to time-domain methods. Patient-specific high-intensity focused ultrasound (HIFU) planning can now incorporate heterogeneous power-law attenuation for transcranial therapy. Quantitative ultrasound (QUS) training datasets can include realistic sub-mm tissue transitions such as tumor margins and fatty infiltration. Aberration correction studies can account for layered abdominal wall structures. The open-source multi-GPU implementation (\hyperlink{https://github.com/pinton-lab/fullwave25}{github.com/pinton-lab/fullwave25}) provides a practical tool for researchers who require high-fidelity tissue models without manual parameter expertise. Future work should validate aberration and reverberation predictions against tissue-mimicking phantoms and extend PML formulations for high-exponent materials ($y > 1.6$). 
\ifarxiv
    \section*{Acknowledgments}
    We would like to thank the University of North Carolina at Chapel Hill and the Research Computing group for providing computational resources and support that have contributed to these research results. Funding provided by NIH R01EB029419 and R01EB036295.

    \section*{Data availability}
    An open-source implementation is available at \url{https://github.com/pinton-lab/fullwave25}.
\else
    \ack{We would like to thank the University of North Carolina at Chapel Hill and the Research Computing group for providing computational resources and support that have contributed to these research results.}
    \funding{Funding provided by NIH R01EB029419 and R01EB036295.}
    \data{Github link: \url{https://github.com/pinton-lab/fullwave25}}
\fi

\bibliographystyle{plainnat}
\bibliography{references}

\ifarxiv
\begin{appendices}

    \begin{figure*}
        \centering
        \includegraphics[width=1.0\linewidth]{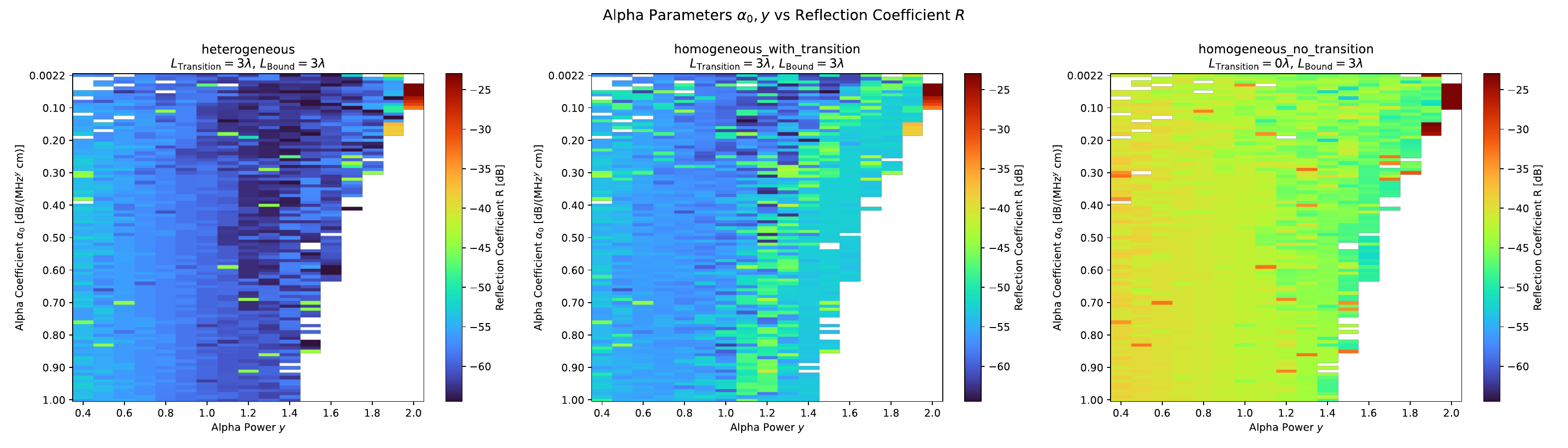}
        \caption{Reflection coefficient vs interior $(\alpha_0, y)$ for three PML configurations, complementing the distribution shown in Fig.~\ref{fig:pml_hom_vs_het}: (a) the adaptive PML used throughout the rest of this work, in which the PML and $3\lambda$ transition layer inherit per-voxel relaxation parameters matched to the interior; (b) a homogeneous PML fixed at $(\alpha_0, y) = (0.5, 1.0)$ with the same $3\lambda$ transition layer; (c) the same homogeneous PML without a transition layer. Lookup-invalid cells (white) are shared across all three panels. The adaptive PML achieves $R < -50$ dB in $97.1\%$ of the $1328$ valid cells; the homogeneous PML with transition layer matches the adaptive baseline near the design point and degrades by up to ${\sim}15$ dB at the corners of the sweep grid; the homogeneous PML without transition layer is uniformly degraded across the grid by ${\sim}16$ dB on the median and triggers numerical instability in $29$ cells.}
        \label{fig:pml_hom_vs_het_heatmap}
    \end{figure*}

\end{appendices}
 \else
    \newpage
    \suppdata{%
\begin{appendices}

    \begin{figure*}
        \centering
        \includegraphics[width=1.0\linewidth]{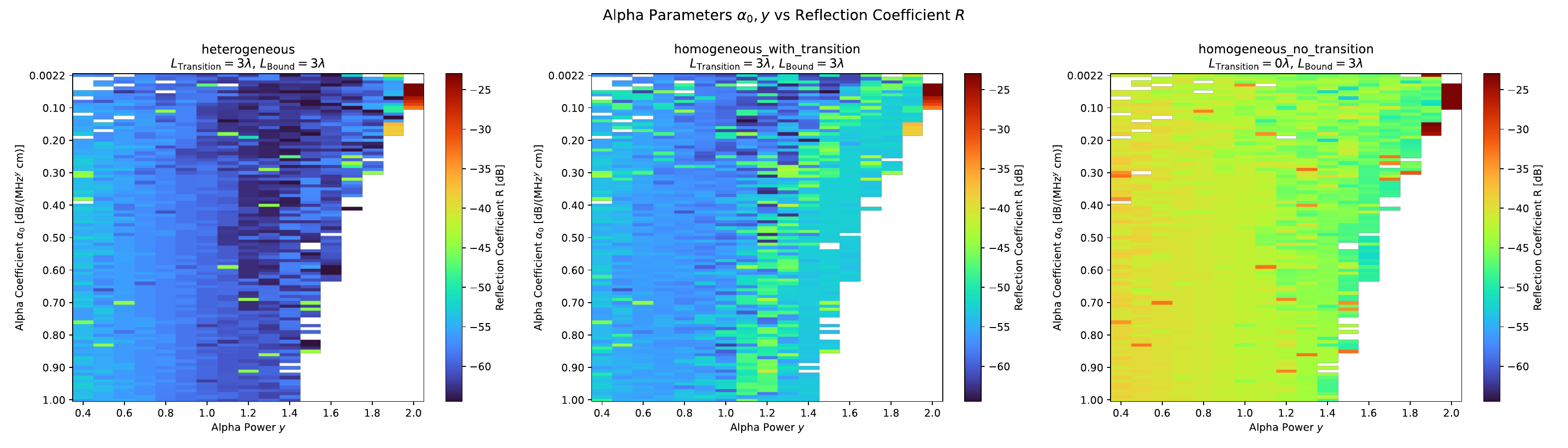}
        \caption{Reflection coefficient vs interior $(\alpha_0, y)$ for three PML configurations, complementing the distribution shown in Fig.~\ref{fig:pml_hom_vs_het}: (a) the adaptive PML used throughout the rest of this work, in which the PML and $3\lambda$ transition layer inherit per-voxel relaxation parameters matched to the interior; (b) a homogeneous PML fixed at $(\alpha_0, y) = (0.5, 1.0)$ with the same $3\lambda$ transition layer; (c) the same homogeneous PML without a transition layer. Lookup-invalid cells (white) are shared across all three panels. The adaptive PML achieves $R < -50$ dB in $97.1\%$ of the $1328$ valid cells; the homogeneous PML with transition layer matches the adaptive baseline near the design point and degrades by up to ${\sim}15$ dB at the corners of the sweep grid; the homogeneous PML without transition layer is uniformly degraded across the grid by ${\sim}16$ dB on the median and triggers numerical instability in $29$ cells.}
        \label{fig:pml_hom_vs_het_heatmap}
    \end{figure*}

\end{appendices}
}
\fi

\end{document}